\documentclass[aps,prb,reprint,amsmath,amssymb,superscriptaddress,nofootinbib,floatfix,showkeys]{revtex4-2}
\usepackage[T1]{fontenc}
\usepackage[utf8]{inputenc}
\usepackage{graphicx}
\usepackage{lmodern}
\usepackage{microtype}
\usepackage{bm}
\usepackage[hidelinks]{hyperref}
\hypersetup{
    colorlinks=true,
    citecolor=blue,
    linkcolor=blue,
    urlcolor=blue
}
\allowdisplaybreaks[1]
\begin{document}

\title{Nonequilibrium energy transport and fluctuations in two-photon-driven nonlinear quantum optical systems}

\author{Yitian Chen}
\affiliation{Department of Physics, Zhejiang Normal University, Jinhua 321004, China}
\author{Yuwei Lu}
\affiliation{Department of Physics, Zhejiang Normal University, Jinhua 321004, China}
\author{Jincheng Lu}
\email{jinchenglu@usts.edu.cn}
\affiliation{Key Laboratory of Intelligent Optoelectronic Devices and Chips of Jiangsu Higher Education Institutions, School of Physical Science and Technology, Suzhou University of Science and Technology, Suzhou 215009, China}
\affiliation{Advanced Technology Research Institute of Taihu Photon Center, School of Physical Science and Technology, Suzhou University of Science and Technology, Suzhou 215009, China}
\author{Chen Wang}
\email{wangchen@zjnu.cn}
\affiliation{Department of Physics, Zhejiang Normal University, Jinhua 321004, China}

\begin{abstract}
Understanding nonequilibrium  transport and  fluctuations 
under a parametric drive in nonlinear quantum optical systems is challenging.
Here, we formulate a driven quantum master equation in the rotating frame combined with full counting statistics, retaining the drive-induced frequency shifts in the system-reservoir interactions and providing a unified description of drive-assisted incoherent transitions based on the rotated system dressed-basis.
Applied to a Kerr resonator and a nonlinear Jaynes-Cummings model, the approach reveals pronounced multiphoton-resonant enhancement of the drive input energy current, with significant high peaks under two-photon driving compared to the single-photon case.
The two-photon resonance relations are analytically obtained.
Resonance structure in the nonlinear Jaynes-Cummings model  nonlinearly relies on qubit-photon couplings, with additional dressed-state branches.
A low-energy-state approximation attributes the resonant current enhancement to dressed-state hybridization and efficient activation of incoherent energy exchange processes.
Beyond the average current, two-photon-driving induced energy exchange picture near resonance also substantially modifies the second-order current fluctuation and higher-order current cumulant, while increasing the time-normalized signal-to-noise ratio.
Adding two-photon loss introduces extra incoherent photon-pair exchange pathways and further reshapes the current and its fluctuations.
We hope these results may
deepen interpretation of photon driving and quantum dissipation cooperatively governing nonequilibrium energy transport and fluctuations.
\end{abstract}


\maketitle
\raggedbottom

\section{Introduction}\label{sec:introduction}

Nonequilibrium energy transport is a central problem in quantum thermodynamics and quantum devices, which concerns energy exchange within a quantum system and between the system and its environment under external driving or a thermodynamic bias~\cite{Fong2019nature,Pekola2021rmp,Landi2022rmp}.
In contrast with classical transport systems, quantum coherence and transition selection rules,  together with dissipative system-environment interactions, reshape energy-transfer channels and control both energy currents and their fluctuations~\cite{Kohler2005pr,esposito2009rmp,Carrega2016prl}.
Current fluctuations are known to characterize the stochastic nature and higher-order statistics of microscopic energy exchange and determine the reliability of current signals in quantum devices~\cite{guarnieri2019prr,Yang2020nc,Tan2022prl,Meier2025np}.
These fluctuations are constrained by thermodynamic uncertainty relations, which impose a trade-off between transport precision and thermodynamic cost~\cite{barato2015prl,gingrich2016prl,Horowitz2020np,Lu2022prb,Wang2024prl}.
This provides a physical basis for understanding the performance limits of quantum transport.

Quantum optical systems nowadays have become crucial platforms for studying nonequilibrium energy transport because of their highly tunable photonic nonlinearities and flexible assembly~\cite{Scully1997book,wallraff2004nature,blais2021rmp,aspelmeyer2014rmp}.
Representative examples include optical Kerr resonators with nonlinear photon interactions~\cite{Lugiato1987prl,Fink2018np,Beaulieu2025nc}, cavity and circuit quantum electrodynamics systems~\cite{wallraff2004nature,blais2021rmp,Clerk2020np} together with lattice extension~\cite{Yarmohammadi2025prb,Yarmohammadi2026prl} including artificial atoms coupled to microwave cavities, and optomechanical systems in which radiation pressure mediates energy exchange between photons and phonons~\cite{aspelmeyer2014rmp,Barzanjeh2022np}.
In these systems, nonlinear photonic interactions compete with quantum dissipation to determine the available exchange processes and  fluctuations of photons and energy currents~\cite{Yang2020nc,Fink2018np,Beaudoin2011pra,Stegmann2022acsp,Ronzani2018np}.
 Quantun noise effect of multiphoton transitions in driven Kerr resonator was explored via the photon number picture~\cite{Leyton2012njp}.
While, nonequilibrium transport fluctuations, based on the energy exchange picture, directly affect the usefulness of flow signals and the amplification performance of quantum devices, and are closely related to their thermodynamic cost~\cite{gingrich2016prl,guarnieri2019prr,koyuk2020prl}.
As is known, nonequilibrium energy transfer through photon exchange has advanced the development of functional quantum devices, including  quantum routers~\cite{hoi2011prl,Zhou2013prl}, quantum rectifiers~\cite{senior2020communphys,Wang2021pra},  quantum amplifiers~\cite{metelmann2015prx},
and quantum thermal transistors~\cite{Majland2020prb}.
Such features are generally exhibited far-from equilibrium, which are apparently distinct from the equilibrium characteristics.
These quantum optical platforms therefore provide natural setups for investigating nonequilibrium energy transport and device effects.

In driven open quantum systems, time-dependent driving provides a key route to realize the far-from-equilibrium state~\cite{Kohler2005pr,koyuk2020prl}. 
Two-photon driving is particularly relevant because pairwise photon creation and annihilation open higher-order transition channels, while generally preserving the parity symmetry of the quantum optical systems~\cite{Deppe2008np}.
Recent superconducting experiments have realized two-photon driving in nonlinear resonators, demonstrating Kerr-cat stabilization and coherent control~\cite{grimm2020nature}, Josephson-based parametric oscillations~\cite{Yamaji2022}, quantum interference in Kerr parametric resonators~\cite{Iyama2024}, and dissipative phase transitions accompanied by squeezing and spontaneous symmetry breaking~\cite{Beaulieu2025nc}.
Circuit-QED experiments have also revealed signatures of photon blockade~\cite{birnbaum2005nature,fink2017prx}.
More broadly, periodic and stochastic driving can reshape energy levels and transition selection rules, thereby controlling energy-exchange channels and directed currents~\cite{Thouless1983prb,Wang2022fop,Segal2008prl,Bai2021advx}, as well as current fluctuations~\cite{Horowitz2020np,Lu2022prb,Takahashi2020prl,Wang2024prl,Gamba2025pra}.
These developments motivate a systematic investigation of nonequilibrium energy transport and its fluctuations under two-photon driving.

From a methodological perspective, weakly coupled open quantum systems are commonly described by Lindblad master equations~\cite{lindblad1976cmp}.
When internal couplings or nonlinearities are appreciable, however, a local description of dissipation may fail to resolve transitions between different dressed states accurately.
By constructing transitions in the system eigenbasis, a dressed master equation can more appropriately account for strong photonic nonlinearities~\cite{Beaudoin2011pra,settineri2018pra}.
Nonequilibrium Green's functions (NEGF) are well suited to continuous spectra and steady-state transport~\cite{Wang2014fop} and have also been extended to transient transport in confined systems~\cite{Tang2014prb}, 
while treating strong nonlinearities and higher-order transport fluctuations within such approaches remains challenging.
The hierarchical equations of motion (HEOM) can incorporate environmental memory effects, but at a substantially higher computational cost~\cite{tanimura1989jpsj,Jin2008jcp}.
For periodically driven systems, Floquet-based NEGF and quantum master equation approaches use quasienergies to describe drive-assisted transitions~\cite{Vahid2024jcp,vahid2025prb}, although their computational cost increases with the Floquet-space expansion.
It is known that time-dependent driving can further modify not only the average energy current but also current noise and thermodynamic uncertainty 
~\cite{Kohler2005pr,Wang2022fop}.
Thus, a computationally efficient and physically concise approach is still demanding to study driven quantum energy transport.
Moreover, how two-photon driving reorganizes nonlinear dressed-state transitions and thereby controls both the average energy current and high-order current fluctuations remains unclear.

Here, we investigate nonequilibrium energy transport and fluctuations in two-photon-driven quantum optical systems by including a driven quantum master equation.
The dissipative transition operators are reconstructed in the eigenbasis of the rotating-frame system Hamiltonian, allowing drive-induced energy-level hybridization and the associated incoherent transition channels to be incorporated consistently. 
We apply this framework to a driven Kerr resonator and a nonlinear Jaynes-Cummings model.
In both systems, multiphoton resonances strongly enhance the driving-induced input energy current, with two-photon driving producing a larger response than single-photon driving.
The two-photon resonance positions are obtained analytically for the Kerr resonator, and the corresponding resonance conditions are extended to the nonlinear Jaynes-Cummings model.
A low-energy dressed-state approximation further yields an analytical expression for the driving-induced input current and relates its resonant enhancement to eigenstate hybridization, population redistribution, and the activation of incoherent transition channels.
Two-photon driving also improves the current signal-to-noise ratio in several resonance regions and substantially modifies the third current cumulant, extending its effects from average transport to high-order non-Gaussian fluctuations.

The remainder of this paper is organized as follows. Section~\ref{sec:methods} introduces the system model, the driven quantum master equation, and full counting statistics. Section~\ref{sec:kerr} examines energy transport, fluctuations, and two-photon loss in the two-photon-driven Kerr resonator and analyzes the mechanism of resonant enhancement. Section~\ref{sec:jcm} discusses the corresponding transport and fluctuation properties of the nonlinear Jaynes-Cummings model. 
We briefly discuss the possible experimental realization in Section ~\ref{sec:realization}.
Section~\ref{sec:conclusions} concludes the paper.

\section{Model and methods}\label{sec:methods}

\subsection{Photon-driven quantum optical systems}

\begin{figure}[thb]
    \centering
    \includegraphics[width=\linewidth]{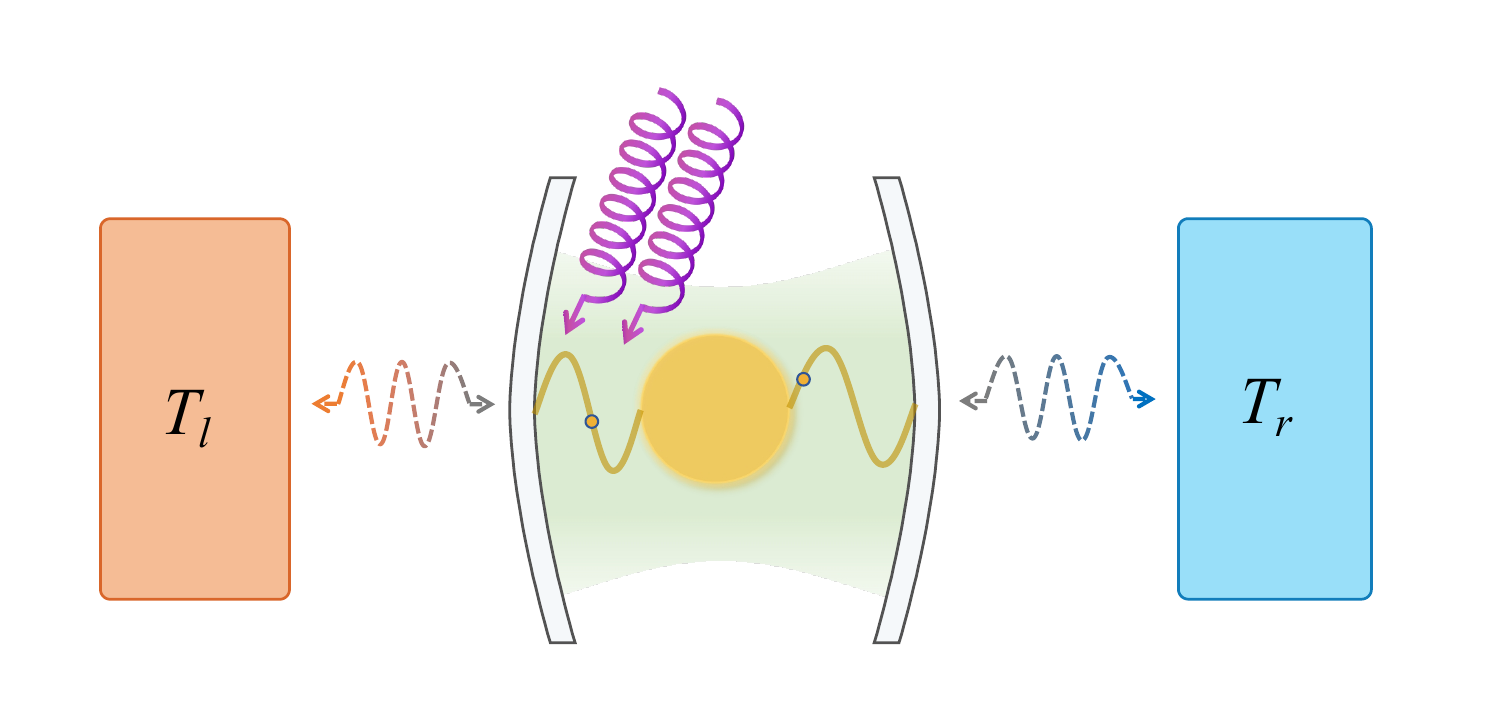}
\caption{Schematic of a nonequilibrium quantum optical system under two-photon driving. The central region represents a driven quantum optical system, such as a light-matter interacting system. The double-helical line with an arrow denotes the external two-photon drive. The bosonic reservoirs on the left and right are at temperatures $T_{l}$ and $T_{r}$, respectively, and the dashed wavy arrows indicate their dissipative couplings to the quantum system.}
    \label{fig:fig1}
\end{figure}

We consider the driven nonequilibrium quantum optical system illustrated in Fig.~\ref{fig:fig1}. An optical cavity embedded with nonlinear quantum matter is coherently driven by an external field and coupled on its left and right to bosonic reservoirs at temperatures $T_{l}$ and $T_{r}$, respectively. When the reservoirs are maintained at different temperatures, $T_{l} \neq T_{r}$, the system--reservoir couplings mediate the exchange of excitations and energy, thereby establishing a nonequilibrium quantum transport configuration.
Under $N_d$-photon driving, the total Hamiltonian is described as (\(\hbar=1\)):
\begin{equation}~\label{drivenH0}
\hat{H}_{\textrm{tot}}(t)=\hat{H}_{\textrm{DS}}(t)+\sum_{\mu=l,r}(\hat{H}_{b,\mu}+\hat{V}_{\mu}).
\end{equation}
The Hamiltonian of the driven photonic system denotes~\label{Deppe2008np}
\begin{equation}~\label{drivenH1}
\hat{H}_{\textrm{DS}}(t)=\hat{H}_{\textrm{S}}-{\eta}[e^{-iN_d\omega_dt}(\hat{a}^\dag)^{N_d}+e^{iN_d\omega_dt}\hat{a}^{N_d}].  
\end{equation}
Here, $\hat{a}^\dag~(\hat{a})$ is the creation (annihilation) operator of the single-mode field, $\hat{H}_{\textrm{S}}(\hat{a},\hat{a}^\dag)$ is the reduced Hamiltonian of the photonic system in the absence of driving, $N_d$ specifies a $N_d$-photon process, and $\eta$ and $\omega_d$ are the driving amplitude and frequency, respectively.
The bosonic reservoirs are expressed as  $\hat{H}_{b,\mu}=\sum_{k}\omega_{k,\mu}\hat{b}^\dag_{k\mu}\hat{b}_{k\mu}$, where $\hat{b}^\dag_{k\mu}~(\hat{b}_{k\mu})$ creates (annihilates) a bosonic excitation of frequency $\omega_{k,\mu}$ in reservoir $\mu$. The linear system-reservoir interaction is described as
\begin{equation}
\hat{V}_\mu=\sum_k(g_{k,\mu}\hat{b}^\dag_{k\mu}\hat{A}_{\mu}+g^{*}_{k,\mu}\hat{b}_{k\mu}\hat{A}_{\mu}^{\dag}),\label{vR0}  
\end{equation}
where $g_{k,\mu}$ is the coupling strength between the system and reservoir $\mu$, and $\hat A_{\mu}$ is the system transition operator coupled to that reservoir. Its explicit form depends on the quantum optical model under consideration.
The system--reservoir coupling is commonly characterized by the spectral function $\gamma_\mu(\omega)=2\pi\sum_k|g_{k,\mu}|^2\delta(\omega-\omega_{k,\mu})$.
Here, we select the representative Ohmic case~\cite{Weiss2008book}, $\gamma_\mu(\omega)=\alpha_\mu\theta(\omega)\omega\exp(-\omega/\omega_c)$, where $\alpha_\mu$ is the dissipation strength, $\omega_c$ is the reservoir cutoff frequency, and $\theta(\omega)$ is the Heaviside step function, with $\theta(\omega>0)=1$ and $\theta(\omega{\leq}0)=0$.
Other forms of the spectral function can also be straightforwardly used.

We next study the dynamics of the driven-dissipative  quantum optical system in the rotating frame.
We apply the transformation $\hat{R}(t)=\exp[-i\omega_dt\hat N]$ to the total density operator, where $\hat N$ is the excitation-number operator appropriate to the model. For the Kerr resonator, $\hat N=\hat a^\dagger\hat a$; for the Jaynes-Cummings model considered below, $\hat N=\hat a^\dagger\hat a+\hat\sigma_+\hat\sigma_-$.
Defining $\hat{\rho}_\textrm{R}(t)=\hat{R}^\dag(t)\hat{\rho}_\textrm{tot}(t)\hat{R}(t)$, we obtain the rotating-frame equation of motion
\begin{equation}~\label{EqR1}
\frac{d}{dt}\hat{\rho}_\textrm{R}(t)=i[\hat{\rho}_\textrm{R}(t),\hat{H}_\textrm{R}(t)].
\end{equation}
The transformed total Hamiltonian is $\hat{H}_\textrm{R}(t)=\hat{H}_{\rm{S},\rm{R}}+\sum_\mu[\hat{H}_{b,\mu}+\hat{V}_{\textrm{R},\mu}(t)]$. The corresponding rotating-frame Hamiltonian of the photonic system is
\begin{equation}\label{eq5}
    \hat{H}_{\rm{S},\rm{R}}=\hat{H}_{\textrm{S}}-\omega_d\hat{N}-{\eta}[(\hat{a}^\dag)^{N_d}+\hat{a}^{N_d}],
\end{equation}
where the static part is assumed to conserve excitation number, $[\hat{N},\hat{H}_{\textrm{S}}]=0$. For brevity, we denote $\hat H\equiv\hat {H}_{\rm{S},\rm{R}}$ below.
The explicitly time-dependent system-reservoir interaction in the rotating frame is
\begin{align}
\hat{V}_{\textrm{R},\mu}(t)&=\sum_k[g_{k,\mu}\hat{b}^\dag_{k\mu}\hat{R}^\dag(t)\hat{A}_\mu\hat{R}(t)+\rm{H.c.}],\label{vR}    
\end{align}

Although the transformation renders the driven-system Hamiltonian in Eq.~(\ref{drivenH1}) time independent, the system--reservoir interaction remains explicitly time dependent.
For example, 
for $\hat A_\mu=\hat a$,
$\hat{V}_{\textrm{R},\mu}(t)=\sum_k(g_{k,\mu}e^{-i\omega_dt}\hat{b}^\dag_{k\mu}\hat{a}+\rm{H.c.})$.
Thus, although the rotating-frame transformation removes the explicit time dependence from the system Hamiltonian, the drive frequency remains encoded in the system--reservoir interaction.

\subsection{Photon-driven quantum master equation}

We now consider the dissipative dynamics of the photonic system under $N_d$-photon driving. In the rotating frame, the system-reservoir interaction is given by Eq.~(\ref{vR}), whose explicitly time-dependent terms retain information about the drive.
We assume weak system--reservoir coupling, such that a perturbative Born--Markov treatment is applicable.
Starting from the interaction-picture von Neumann equation $\frac{d}{dt}\hat{\rho}^\textrm{R}_{I}(t)=-i\sum_\mu[\hat{V}^\textrm{R}_{\mu,I}(t),\hat{\rho}^\textrm{R}_{I}(t)]$, we define $\hat{\rho}^\textrm{R}_{I}(t)=\exp(i\hat{H}_0t)\hat{\rho}^\textrm{R}(t)\exp(-i\hat{H}_0t)$, with $\hat{H}_0=\hat{H}_{\rm{S},\rm{R}}+\sum_{\mu}\hat{H}_{b,\mu}$.
Under weak system--reservoir coupling, the Born approximation factorizes the total density operator as $\hat{\rho}^\textrm{R}_{I}(t){\approx}\hat{\rho}^\textrm{R}_{s,I}(t){\otimes}[\Pi_{\mu}\hat{\rho}_{b,\mu}]$, where the equilibrium reservoir density operator is $\rho_{b,\mu}=\exp(-\beta_{\mu}\hat{H}_{b,\mu})/\textrm{Tr}[\exp(-\beta_{\mu}\hat{H}_{b,\mu})]$.
Assuming that the reservoir correlation time is short compared with the characteristic system timescale, we further apply the Markov approximation, replacing $\hat{\rho}^\textrm{R}_{s,I}(\tau){\approx}\hat{\rho}^\textrm{R}_{s,I}(t)$, and tracing out the bosonic reservoirs yields the equation of motion for the reduced system density operator,
\begin{equation}~\label{append:qme1}
 \frac{d\hat{\rho}^\textrm{R}_{s,I}(t)}{dt}= -\sum_{\mu}\int^t_0d\tau\textrm{Tr}_b\{
 [\hat{V}^\textrm{R}_{\mu,I}(t),[\hat{V}^\textrm{R}_{\mu,I}(\tau),\hat{\rho}^\textrm{R}_{s,I}(t)\otimes
\Pi_{\nu}\hat{\rho}_{b,\nu}]]\}.
 \end{equation}
A key step is to obtain the explicit form of $\hat{V}^\textrm{R}_{\mu,I}(t)=e^{i\hat{H}_0t}\hat{V}^\textrm{R}_{\mu}(t)e^{-i\hat{H}_0t}$, namely,
\begin{equation}
    \hat{V}^\textrm{R}_{\mu,I}(t)=\sum_k[g_{k,\mu}e^{i(\omega_{k,\mu}-\omega_d)t}\hat{b}^\dag_{k,\mu}\hat{A}_{\mu,I}(t)
+\rm{H.c.}],
\end{equation}
where $\hat{A}_{\mu,I}(t)=e^{i\hat{H}t}\hat{A}_{\mu}e^{-i\hat{H}t}$. Within the Markov approximation, we extend the upper limit of the memory integral to infinity and change the integration variable to the time delay $\tau$. Transforming Eq.~(\ref{append:qme1}) back to the Schr\"{o}dinger picture and keeping terms to second order in the system--reservoir coupling yields the driven quantum master equation (dQME),
\begin{eqnarray}
d\hat{\rho}_{s,\rm R}(t)/dt
&=&i[\hat{\rho}_{s,\rm R},\hat{H}_{\rm{S},\rm{R}}]+\sum_\mu\{([\hat{D}^\dag_{\mu,+}\hat{\rho}_{s,\rm R},\hat{A}_\mu]\nonumber\\
&&+[\hat{D}_{\mu,-}\hat{\rho}_{s,\rm R},\hat{A}^\dag_\mu])+\rm H.c.\},\label{qme1}
\end{eqnarray}
Here, $\hat{\rho}_{s,\rm R}(t)=\hat{R}^\dag(t)\hat{\rho}_{s}(t)\hat{R}(t)$ is the system density operator in the rotating frame, and the modified system operators are
\begin{subequations}
	\begin{align}
\hat{D}_{\mu,+}&=\int^\infty_0d\tau\sum_k|g_{k,\mu}|^2n_{k,\mu} e^{-i(\omega_{k,\mu}-\omega_d)\tau}\hat{A}_\mu(-\tau),~\label{Dmu+}\\
\hat{D}_{\mu,-}&=\int^\infty_0d\tau\sum_k|g_{k,\mu}|^2(1+n_{k,\mu}) e^{-i(\omega_{k,\mu}-\omega_d)\tau}\hat{A}_\mu(-\tau),~\label{Dmu-}
\end{align}
\end{subequations}
where $n_{k,\mu}=1/[\exp(\omega_{k,\mu}/k_{\textrm{B}}T_\mu)-1]$ is the Bose--Einstein distribution and $\hat{A}_\mu(-\tau)=e^{-i\hat{H}_{\rm S,R}\tau}\hat{A}_{\mu}e^{i\hat{H}_{\rm S,R}\tau}$ is the system operator.
A key feature of the present rotating-frame formulation is that the kernels defining $\hat D_{\mu,\pm}$ retain an explicit
drive-frequency-dependent phase factor. Consequently, the reservoir-assisted transition rates depend explicitly on the drive frequency.

\subsection{Driven quantum master equation with full counting statistics}

Full counting statistics (FCS) provides a systematic framework for characterizing current fluctuations in nonequilibrium systems~\cite{esposito2009rmp}. Beyond the average current, FCS systematically captures fluctuations and noise in steady-state transport through a cumulant-generating function, making it particularly suitable for analyzing energy exchange in driven open quantum systems.
Consider a general driven nonequilibrium quantum system,
$\hat{H}_{\textrm{tot}}(t)=\hat{H}_{\textrm{DS}}(t)+\sum_{\mu}(\hat{H}_{b,\mu}+\hat{V}_\mu)$.
To characterize the statistics of energy exchanged with the reservoirs, we introduce the counting fields $\chi_\mu$ through
$\hat{H}^\chi_{\textrm{tot}}(t)=e^{i\sum_\mu\chi_\mu\hat{H}_{b,\mu}/2}\hat{H}_{\textrm{tot}}(t)e^{-i\sum_\mu\chi_\mu\hat{H}_{b,\mu}/2}$.
The resulting Hamiltonian is
$\hat{H}^\chi_{\textrm{tot}}(t)=\hat{H}_{\textrm{DS}}(t)+\sum_{\mu}(\hat{H}_{b,\mu}+\hat{V}_{\chi_\mu})$, where the counting-field-dependent system--reservoir interaction is
\begin{eqnarray}
\hat{V}_{\chi_\mu}=\sum_k(g_{k,\mu}e^{i\omega_{k,\mu}\chi_\mu/2}\hat{b}^\dag_{k,\mu}\hat{A}_\mu+\rm{H.c.}).\label{vR0chi}
\end{eqnarray}
The corresponding counting-field-dependent density operator is defined as 
\begin{eqnarray}
    \hat{\rho}^{\chi}_{\rm tot}(t)=\hat{U}_{\chi}(t,0)\hat{\rho}_{\rm tot}(0)\hat{U}^{\dagger}_{-\chi}(t,0),
\end{eqnarray}
where \(\hat{U}_{\pm\chi}(t,0)\) are generated by \(\hat{H}^{\pm\chi}_{\rm tot}(t)\).
After the rotating-frame transformation $\hat{\rho}^\chi_\textrm{R}(t)=\hat{R}^\dag(t)\hat{\rho}^\chi_\textrm{tot}(t)\hat{R}(t)$, the counting-field-dependent density operator obeys the tilted von Neumann equation $\frac{d}{dt}\hat{\rho}^\chi_\textrm{R}(t)=-i[\hat{H}^\chi_\textrm{R}(t)\hat{\rho}^\chi_\textrm{R}(t)-\hat{\rho}^\chi_\textrm{R}(t)\hat{H}^{-\chi}_\textrm{R}(t)]$.
Here, $\hat{H}^{\pm\chi}_\textrm{R}(t)=\hat{H}_{\rm S,R}+\sum_{\mu}(\hat{H}_{b,\mu}+\hat{V}^\textrm{R}_{\pm\chi_\mu}(t))$, with the modified interaction
$\hat{V}^\textrm{R}_{\chi_\mu}(t)=\sum_k(g_{k,\mu}e^{-i\omega_dt}e^{i\omega_{k,\mu}\chi_\mu/2}\hat{b}^\dag_{k,\mu}\hat{A}_\mu+\rm{H.c.})$. The interaction $\hat V^\textrm{R}_{-\chi_\mu}(t)$ is obtained by replacing $\chi_\mu\rightarrow-\chi_\mu$.
Within the Born-Markov approximation, the dQME incorporating FCS is
\begin{eqnarray}~\label{DQMEFCS}
\frac{d}{dt}\hat{\rho}^\chi_{{s,\rm{R}}}(t)
&=&i[\hat{\rho}^\chi_{{s,\rm{R}}},\hat{H}_{\rm{S},\rm{R}}]\\
&&-\sum_\mu\{(\hat{A}_\mu\hat{D}^\dag_{\mu,+}\hat{\rho}^\chi_{s,\rm{R}}+\hat{A}^\dag_\mu\hat{D}_{\mu,-}\hat{\rho}^\chi_{s,\rm{R}})\nonumber\\
&&\quad+(\hat{\rho}^\chi_{{s,\rm{R}}}\hat{D}_{\mu,+}\hat{A}^\dag_\mu+\hat{\rho}^\chi_{{s,\rm{R}}}\hat{D}^\dag_{\mu,-}\hat{A}_\mu)\nonumber\\
&&\quad-(\hat{A}_\mu\hat{\rho}^\chi_{{s,\rm{R}}}\hat{D}^\dag_{\mu,-}(\chi_\mu)+\hat{A}^\dag_\mu\hat{\rho}^\chi_{_{\rm{S},\rm{R}}}\hat{D}_{\mu,+}(\chi_\mu))\nonumber\\
&&\quad-(\hat{D}_{\mu,-}(-\chi_\mu)\hat{\rho}^\chi_{{s,\rm{R}}}\hat{A}^\dag_\mu+\hat{D}^\dag_{\mu,+}(-\chi_\mu)\hat{\rho}^\chi_{{s,\rm{R}}}\hat{A}_\mu)\},\nonumber
\end{eqnarray}
where modified system operators are given by
\begin{subequations}
 	\begin{align}
\hat{D}_{\mu,+}(\chi_\mu)&=\sum_{n,m}\frac{\gamma_\mu(\omega_d+E_{mn})}{2}n_\mu(\omega_d+E_{mn})\nonumber\\*&\quad\times e^{-i(\omega_d+E_{mn})\chi_\mu}A^{nm}_{\mu}|\varphi_n{\rangle}{\langle}\varphi_m|,~\label{Dpchi}\\
\hat{D}_{\mu,-}(\chi_\mu)&=\sum_{n,m}\frac{\gamma_\mu(\omega_d+E_{mn})}{2}[1+n_\mu(\omega_d+E_{mn})]\nonumber\\*&\quad\times e^{-i(\omega_d+E_{mn})\chi_\mu}A^{nm}_{\mu}|\varphi_n{\rangle}{\langle}\varphi_m|~\label{Dmchi}.
\end{align}
\end{subequations}
Here, $\hat H|\phi_n\rangle=E_n|\phi_n\rangle$,
$E_{mn}=E_m-E_n$, and
$A^{nm}_{\mu}=\langle\phi_n|\hat A_\mu|\phi_m\rangle$.
The spectral function is $\gamma_\mu(\omega)=2\pi\sum_k|g_{k,\mu}|^2\delta(\omega-\omega_{k,\mu})$, and the Bose-Einstein distribution is $n_\mu(\omega)=1/[\exp(\omega/k_{\textrm{B}}T_\mu)-1]$.
Once all $\chi_\mu=0$, Eq.~(\ref{DQMEFCS}) is completely reduced to Eq.~(\ref{qme1}).

In the long-time limit, the cumulant-generating function per unit time is $\mathcal{G}(\chi)=\lim_{t\rightarrow{\infty}}\ln{[\sum_{\alpha}\rho^\chi_{\alpha\alpha}(t)]}/t$,
with $\rho^\chi_{\alpha\alpha}(t)={\langle}\phi_\alpha|\hat{\rho}^\chi_{s,\rm R}(t)|\phi_\alpha{\rangle}$.
The energy current into reservoir $\mu$ and its fluctuations follow from $J^{(n)}_\mu=\frac{{\partial}^n\mathcal{G}(\chi)}{{\partial}(i\chi_\mu)^n}\Big|_{\chi=0}$.
In particular, the steady-state energy current, second current cumulant, and third current cumulant for reservoir $\mu$ are, respectively,
\begin{subequations}
\begin{align}
    J_\mu=&\frac{{\partial}\mathcal{G}(\chi)}{{\partial}(i\chi_\mu)}\Big|_{\chi=0},\\
     J^{(2)}_\mu=&\frac{{\partial}^2\mathcal{G}(\chi)}{{\partial}(i\chi_\mu)^2}\Big|_{\chi=0},\\
         J^{(3)}_\mu=&\frac{{\partial}^3\mathcal{G}(\chi)}{{\partial}(i\chi_\mu)^3}\Big|_{\chi=0}.
\end{align}
\end{subequations}
We adopt the convention that $J_\mu>0$ denotes energy flowing into reservoir $\mu$. 
At steady state, energy conservation indirectly gives the energy current as $J_{\rm P}=J_l+J_r$ from the driving port.

\section{Two-photon-driven Kerr resonator}\label{sec:kerr}
As a representative quantum optical system, the Kerr resonator has been extensively studied in photon-correlation measurements~\cite{Fink2018np}, quantum criticality~\cite{Krimer2019prl,Beaulieu2025nc}, quantum sensing~\cite{Heugel2019prl}, and quantum state preparation~\cite{Puri2017npj}, among other areas.
It also provides a standard effective model for nonlinear photonic interactions, with the system Hamiltonian
$\hat{H}_{\rm{S}}=\omega_0 \hat{a}^{\dagger}\hat{a}
+U\hat{a}^{\dagger}\hat{a}^{\dagger}\hat{a}\hat{a}$,
where $\omega_0$ is the resonator frequency and $U$ is the Kerr nonlinearity.
For a two-photon drive, i.e., $-\eta[e^{2i\omega_dt}\hat{a}^2+H.c.]$, Eq.~(\ref{eq5}) gives the rotating-frame system Hamiltonian
\begin{equation}~\label{HKerrR1}
\hat{H}_{\rm{Kerr}}
=\Delta \hat{a}^{\dagger}\hat{a}
+U\hat{a}^{\dagger}\hat{a}^{\dagger}\hat{a}\hat{a}
-\eta\left[(\hat{a}^{\dagger})^2+\hat{a}^2\right],
\end{equation}
where $\Delta=\omega_0-\omega_d$ is the detuning between the cavity mode and the drive. This Hamiltonian shows that the external two-photon drive creates or annihilates photons in pairs, thereby modifying the system eigenstates and the conditions for transitions between them.
Under nonequilibrium conditions, $T_{l}\neq T_{r}$, the system exchanges energy incoherently with the left and right reservoirs. The combined action of periodic driving and system--environment dissipation therefore generates a steady-state energy current.

\subsection{Energy current under two-photon driving}

\begin{figure}[tb]
    \centering
    \includegraphics[width=\linewidth]{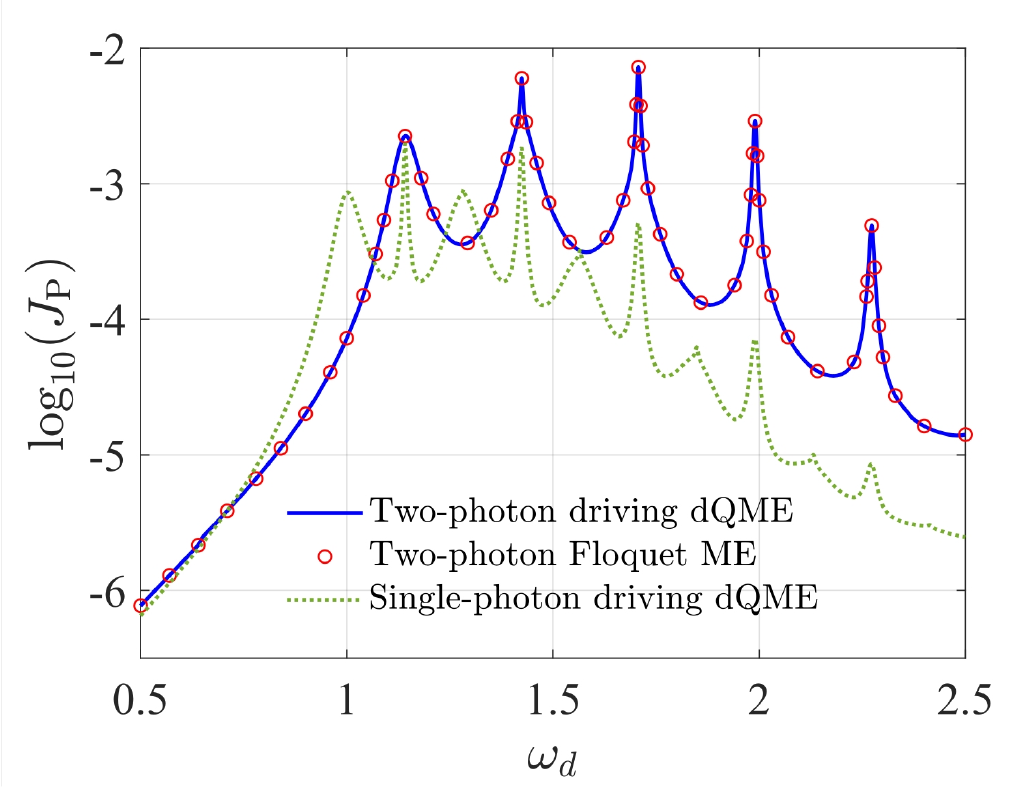}
\caption{Comparison of the drive input energy current in the Kerr resonator as a function of driving frequency for different driving protocols. The blue solid line shows the result of the two-photon-driven quantum master equation, the red circles show the Floquet master equation result, and the green dashed line shows the result of the single-photon-driven quantum master equation. The parameters are the photon-number cutoff $N_{\rm{ph}}=30$, resonator frequency $\omega_0=1$, driving amplitude $\eta=0.02\omega_0$, Kerr nonlinearity $U=\omega_0/(5\sqrt{2})$, dissipation strengths $\alpha_{l}=\alpha_{r}=0.001$, cutoff frequency $\omega_c=10\omega_0$, and reservoir temperatures $k_{\rm B}T_{l}=0.8\omega_0$ and $k_{\rm B}T_r=0.4\omega_0$.}
    \label{fig:fig2}
\end{figure}

We first examine the steady-state drive input energy current $J_{\rm{P}}$ as a function of driving frequency $\omega_d$, as shown in Fig.~\ref{fig:fig2}.
At weak driving amplitudes, the current obtained from the two-photon-driven quantum master equation agrees closely with the Floquet master equation result over the frequency range shown.
In particular, the two approaches yield the same peak structure near resonance.
This agreement confirms the accuracy of the driven quantum master equation for steady-state transport in the two-photon-driven Kerr resonator within the weak-driving regime considered here.
A comparison of two-photon and single-photon driving reveals clear differences in both the resonance positions and the peak currents, with two-photon driving producing larger peaks in several resonance regions. We focus below on nonequilibrium energy transport under two-photon driving and its physical mechanism.

\begin{figure*}[t]
    \centering
    \includegraphics[width=\linewidth]{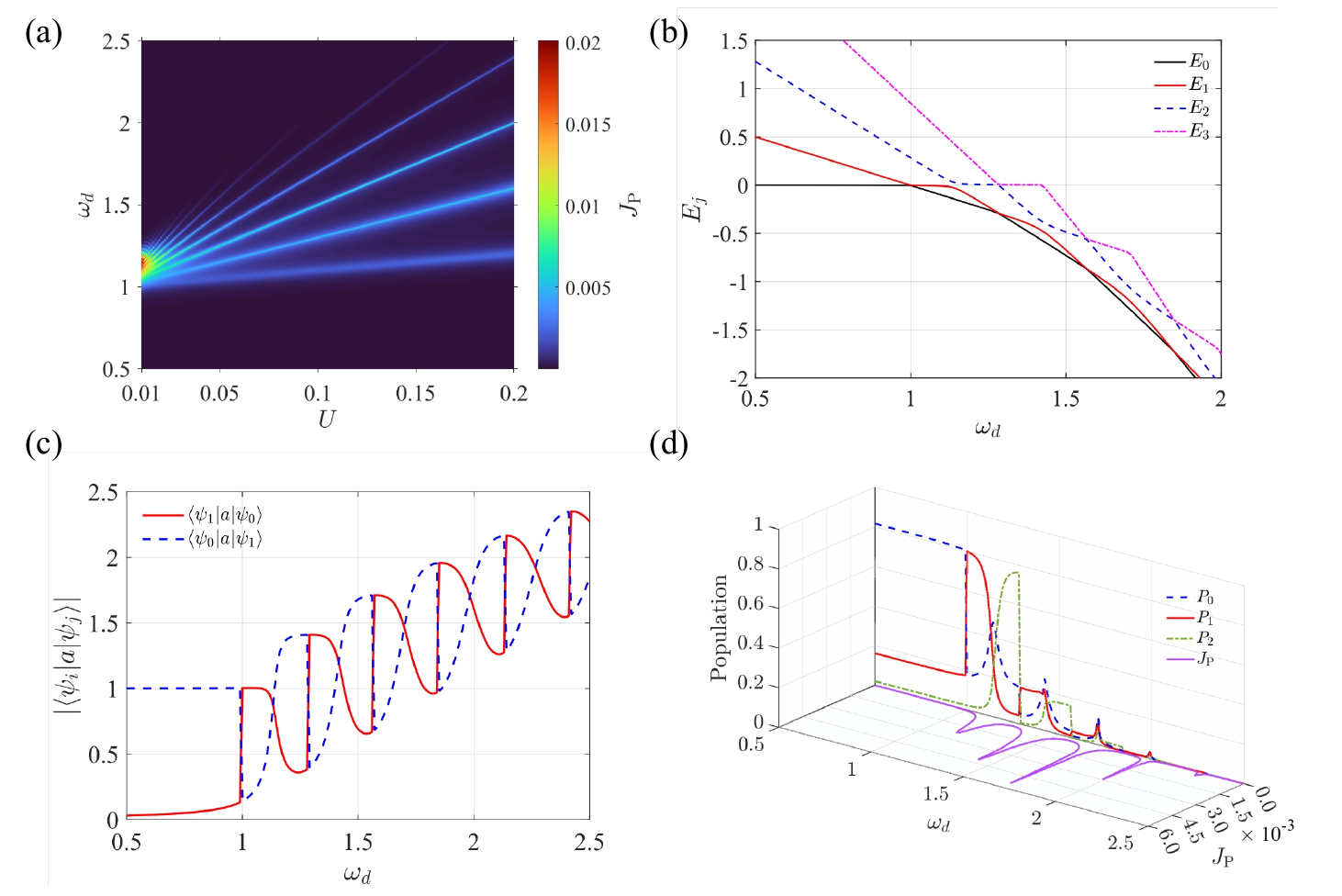}
\caption{Resonance structure and physical mechanism of the drive input energy current in the two-photon-driven Kerr resonator. (a) Top view of the three-dimensional distribution of $J_{\rm{P}}$ as a function of Kerr nonlinearity $U$ and driving frequency $\omega_d$.
(b) Dependence of the system energy-level spacings $\Delta$ on driving frequency $\omega_d$. (c) Relevant transition matrix elements as functions of $\omega_d$. (d) Low-energy-state populations $P_0$, $P_1$, and $P_2$, together with $J_{\rm{P}}$, as functions of $\omega_d$. In (a), higher-order resonances associated with two-photon transitions are confined to very narrow parameter regions at large $U$, causing the numerically obtained resonance lines to appear approximately discrete. For panels (b)-(d), $U=\omega_0/(5\sqrt{2})$. The remaining parameters are $N_{\rm{ph}}=30$, $\omega_0=1$, $\eta=0.02\omega_0$, $\alpha_{l}=\alpha_{r}=0.001$, $\omega_c=10\omega_0$, $k_{\rm B}T_{l}=0.8\omega_0$, and $k_{\rm B}T_{r}=0.4\omega_0$.}
    \label{fig:fig3}
\end{figure*}

Figure~\ref{fig:fig3}(a) shows the influence of driving frequency and Kerr nonlinearity on the drive input energy current $J_{\rm P}$.
The current is strongly enhanced along several bright resonance lines.
As $U$ increases, the Kerr nonlinearity produces increasingly unequal photon energy-level spacings, causing the resonance lines to shift appreciably.
This suggests that the current under two-photon driving depends strongly on the Kerr energy-level structure and 
is strongly enhanced when twice the driving frequency approaches the relevant two-photon transition energy.
The undriven Kerr Hamiltonian $\hat{H}_{\rm{S}}=\omega_0 \hat{a}^{\dagger}\hat{a}
+U\hat{a}^{\dagger}\hat{a}^{\dagger}\hat{a}\hat{a}$ gives the energy of the $m$th Fock state as
$\mathcal{E}_m=m\omega_0+Um(m-1)$.
The two-photon drive term $[(\hat{a}^{\dagger})^2+\hat{a}^2]$ induces coherent transitions between $|m-1\rangle$ and $|m+1\rangle$, whose energy difference is
$\mathcal{E}_{m+1,m-1}=\mathcal{E}_{m+1}-\mathcal{E}_{m-1}=2\omega_0+2(2m-1)U$.
Since each two-photon driving process exchanges two drive photons of frequency $\omega_d$, the resonance condition in the weak-driving regime is approximately
$2\omega_d{\approx}\mathcal{E}_{m+1,m-1}$.
The position of the $m$th drive-induced two-photon resonance is therefore
\begin{equation}~\label{resonance1}
\omega_d^{(m)}
{\approx}\omega_0+(2m-1)U,
\qquad m=1,2,\ldots .
\end{equation}
As the Kerr nonlinearity $U$ increases, the two-photon resonance frequencies shift toward higher frequencies, with higher-order resonances moving at larger slopes. This produces the discrete resonance branches extending from low to high frequencies in Fig.~\ref{fig:fig3}(a).
Even a weak drive further hybridizes the Kerr-resonator energy levels, so the actual peak positions can deviate slightly from these estimates.
Increasing the Kerr nonlinearity $U$ also raises the transition energies associated with highly excited states, substantially suppressing their thermal populations at finite temperatures.
In the weak-driving regime, drive-induced resonance broadening is particularly small.
Consequently, higher-order two-photon transitions are efficiently activated only in narrow parameter regions where $2\omega_d$ closely matches the corresponding energy difference, producing the approximately discrete higher-order resonance branches, as shown in Fig.~\ref{fig:fig3}(a).

To clarify the physical mechanism underlying the current peaks, we try to give an approximate expression of  the drive input current in the rotating frame.
For the parameter regime considered here, the off-diagonal elements of the density matrix in the system eigenbasis are numerically negligible, indicating no appreciable bath-induced coherence~\cite{Li2015aop,Zou2024npj}.
At low temperatures, it is found that retaining the three lowest energy states already captures the resonance feature.
Hence, we include three lowest energy states to approximately obtain the drive input energy current, whose validity, though not shown here, has also been verified numerically (see Appendix~\ref{app:currents}):
\begin{eqnarray}
    J_{\rm P}&\approx&\sum_{j=1,2}\{ (\omega_d+\Lambda_j)\times\nonumber\\
&&\quad[\Gamma_-(\omega_d+\Lambda_j)P_j-\Gamma_+(\omega_d+\Lambda_j)P_0]\nonumber\\
&&+(\omega_d-\Lambda_j)\times\nonumber\\
&&\quad[\Gamma_-(\omega_d-\Lambda_j)P_0-\Gamma_+(\omega_d-\Lambda_j)P_j]\}.\label{Jp3TL}
\end{eqnarray}
Here, $\Lambda_j=E_{j,m}-E_{0,m}$ are the two excitation gaps, $P_j~(j=0,1,2)$ are the steady-state populations of the three lowest levels, and $\Gamma_{\pm}(\omega)=\sum_{\mu=l,r}\Gamma^\mu_{\pm}(\omega)$ are the total incoherent transition rates.
The reservoir-assisted transition rates are
\begin{subequations}
	\begin{align}
\Gamma^\mu_{-}(\omega_d+E_{m^{\prime}m})&=\gamma_\mu(\omega_d+E_{m^{\prime}m})[1+n_\mu(\omega_d+E_{m^{\prime}m})]\nonumber\\*&\quad \times|{\langle}\phi_m|\hat{a}|\phi_{m^\prime}{\rangle}|^2,~\label{Rmu-}\\
\Gamma^\mu_{+}(\omega_d+E_{m^{\prime}m})&=\gamma_\mu(\omega_d+E_{m^{\prime}m})n_\mu(\omega_d+E_{m^{\prime}m})\nonumber\\*&\quad \times|{\langle}\phi_m|\hat{a}|\phi_{m^\prime}{\rangle}|^2.~\label{Rmu+}
\end{align}
\end{subequations}
The first term on the right-hand side of Eq.~(\ref{Jp3TL}) describes drive-assisted channels in which the system and reservoirs exchange an energy $\hbar(\omega_d+\Lambda_j)$.
Here, $\Gamma_-(\omega_d+\Lambda_j)$ describes relaxation from the excited state $|\phi_j\rangle$ to the ground state $|\phi_0\rangle$, releasing energy into the reservoirs, whereas $\Gamma_+(\omega_d+\Lambda_j)$ describes the reverse thermal excitation process.
Their transition strengths are determined by the matrix element $|\langle\phi_0|\hat{a}|\phi_j\rangle|^2$.
The second term describes another branch of drive-assisted incoherent exchange channels with energy $\hbar(\omega_d-\Lambda_j)$.
In this branch, $\Gamma_-(\omega_d-\Lambda_j)$ and $\Gamma_+(\omega_d-\Lambda_j)$ correspond to excitation and relaxation processes, respectively, with transition strengths determined by $|\langle\phi_j|\hat{a}|\phi_0\rangle|^2$.
Equation~(\ref{Jp3TL}) is a key analytical result of this work.

For weak driving and $\omega_d>\omega_0$, the $m$th resonance of the driven Kerr Hamiltonian $\hat{H}_{\rm Kerr}$ in Eq.~(\ref{HKerrR1}) occurs between $|m+1{\rangle}$ and $|m-1{\rangle}$, at the position given by Eq.~(\ref{resonance1}).
Near resonance, the split energy levels are approximately
$E_{2,m}{\approx}(\omega_0-\omega_d)(m+1)+U(m+1)m+\eta\sqrt{m(m+1)}$
and
$E_{1,m}{\approx}(\omega_0-\omega_d)(m-1)+U(m-1)(m-2)-\eta\sqrt{m(m+1)}$,
while the ground-state energy is $E_{0,m}{\approx}(\omega_0-\omega_d)m+Um(m-1)$, as shown in Fig.~\ref{fig:fig3}(b).
The corresponding second excited state, first excited state, and ground state of $\hat{H}_{\rm Kerr}$ are
\begin{subequations}
    \begin{align}
    |\phi_{2,m}{\rangle}{\approx}&(|m+1{\rangle}-|m-1{\rangle})/\sqrt{2},~\label{p2r}\\
|\phi_{1,m}{\rangle}{\approx}&(|m+1{\rangle}+|m-1{\rangle})/\sqrt{2},~\label{p1r}\\
|\phi_{0,m}{\rangle}{\approx}&|m{\rangle}~\label{p0r}.
    \end{align}
\end{subequations}
Intriguingly, the incoherent transition coefficients in Eqs.~(\ref{Rmu-}) and (\ref{Rmu+}) are specified as
\begin{subequations}
\begin{align}
|{\langle}\phi_{j,m}|\hat{a}|\phi_{0,m}{\rangle}|
{\approx}&\sqrt{m/2},\\
|{\langle}\phi_{0,m}|\hat{a}|\phi_{j,m}{\rangle}|
{\approx}&\sqrt{(m+1)/2}~(j=1,2).
\end{align}
\end{subequations}
All incoherent energy-exchange channels between $|\phi_{0,m}{\rangle}$ and $|\phi_{j,m}{\rangle}$ are thus effectively opened,
as exemplified in Fig.~\ref{fig:fig3}(c).
Thus, the corresponding rates $\Gamma^\mu_{\pm}(\omega_d{\pm}\Lambda_j)$ become nonzero.
Meanwhile, the excited-state populations $P_{1}$ and $P_2$ increase substantially, as shown in Fig.~\ref{fig:fig3}(d).
These components  actively contribute to enhancement of the steady-state current (\ref{Jp3TL}).

However, away from the $m$th resonance in $\omega_d$, the eigenstates of $\hat{H}_{\rm Kerr}$ approach photon Fock states.
For example, below the $m$th resonance,
$|\phi_{2,m}{\rangle}{\approx}|m+1{\rangle}$,
$|\phi_{1,m}{\rangle}{\approx}|m-1{\rangle}$,
and $|\phi_{0,m}{\rangle}{\approx}|m{\rangle}$.
The relevant matrix elements then become
${\langle}\phi_{0,m}|\hat{a}|\phi_{2,m}{\rangle}
{\approx}\sqrt{m+1}$ and
${\langle}\phi_{1,m}|\hat{a}|\phi_0{\rangle}
{\approx}\sqrt{m}$.
The drive input current reduces to
\begin{eqnarray}
J_{\rm P}&\approx&(\omega_d+\Lambda_2)[\Gamma_-(\omega_d+\Lambda_2)P_2-\Gamma_+(\omega_d+\Lambda_2)P_0]
\nonumber\\&&+ (\omega_d-\Lambda_1)[\Gamma_-(\omega_d-\Lambda_1)P_0-\Gamma_+(\omega_d-\Lambda_1)P_1]. \nonumber\\
\end{eqnarray}
Accordingly, some of the incoherent transport channels connecting $|\phi_0{\rangle}$ and $|\phi_j{\rangle}$ are closed, directly suppressing the energy current.
Therefore, we conclude that the resonance-induced hybridization of photonic Fock state at  Eqs.~(\ref{p2r}-\ref{p0r}) establishes efficiently incoherent energy exchange processes, strongly enhancing the drive input energy current.


\begin{figure*}[t]
    \centering
    \includegraphics[width=\linewidth]{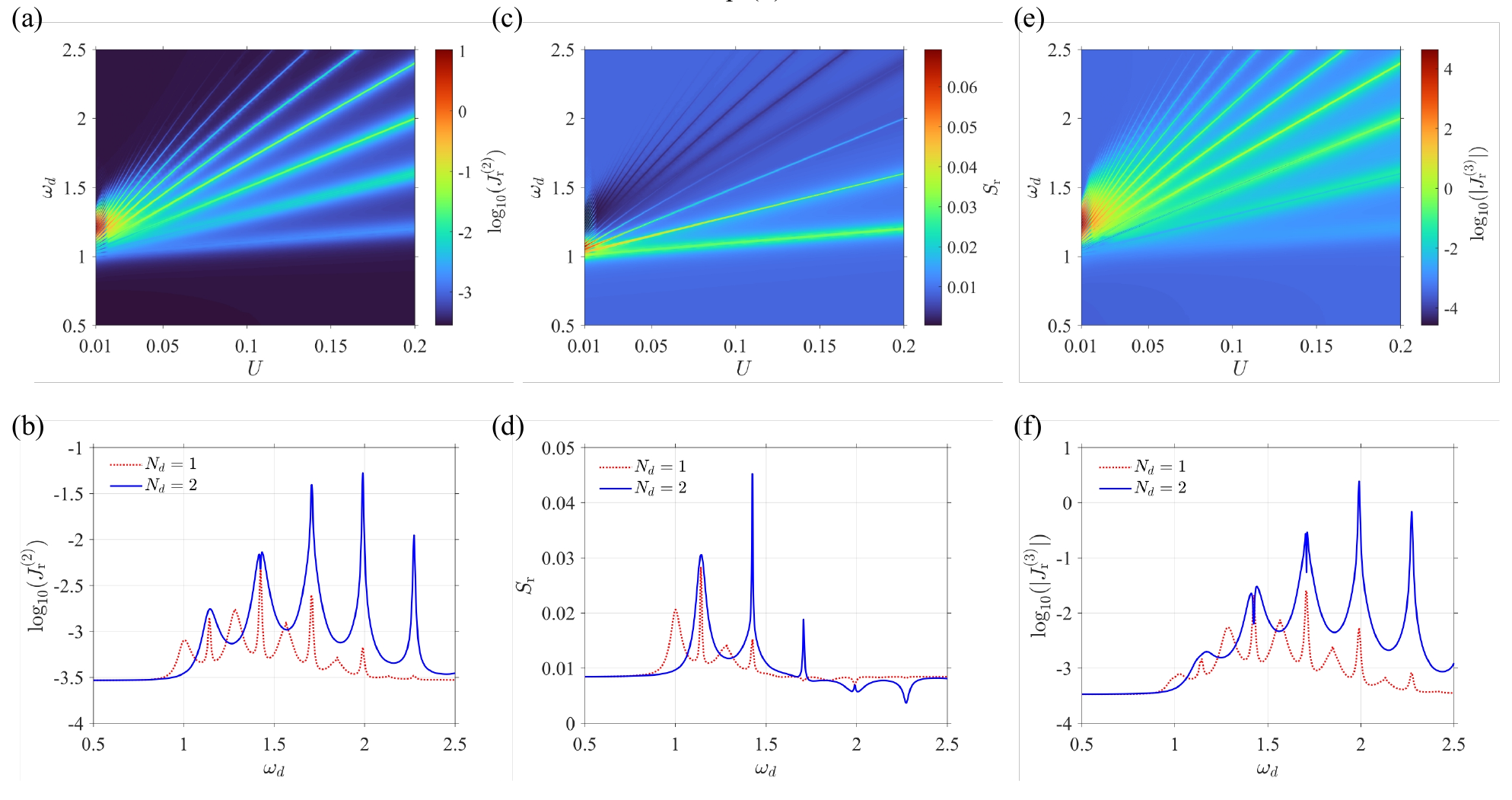}
\caption{Parameter dependence of energy-current fluctuations, the signal-to-noise ratio, and the third current cumulant in the photon-driven Kerr resonator. (a) Top view of the three-dimensional distribution of the second-order current fluctuations $\log_{10}(J^{(2)}_{\rm{r}})$ as a function of Kerr nonlinearity $U$ and driving frequency $\omega_d$. (b) $\log_{10}(J^{(2)}_{\rm{r}})$ as a function of $\omega_d$ at $U=\omega_0/(5\sqrt{2})$. (c) Top view of the three-dimensional distribution of the signal-to-noise ratio $\mathcal{S}_r$ as a function of $U$ and $\omega_d$. (d) $\mathcal{S}_r$ as a function of $\omega_d$ at $U=\omega_0/(5\sqrt{2})$. (e) Top view of the three-dimensional distribution of the third current cumulant $\log_{10}(|J^{(3)}_{\rm r}|)$ as a function of $U$ and $\omega_d$. (f) Two-dimensional section of $\log_{10}(|J^{(3)}_{\rm r}|)$ at $U=\omega_0/(5\sqrt{2})$. In (b), (d), and (f), the red dashed and blue solid lines represent single-photon and two-photon driving, respectively. For panels (b), (d), and (f), $U=\omega_0/(5\sqrt{2})$. The remaining parameters are $N_{\rm ph}=30$, $\omega_0=1$, $\eta=0.02\omega_0$, $\alpha_{l}=\alpha_{r}=0.001$, $\omega_c=10\omega_0$, $k_{\rm{B}}T_{l}=0.8\omega_0$, and $k_{\rm{B}}T_{r}=0.4\omega_0$.}
    \label{fig:fig4}
\end{figure*}

\subsection{Steady-state current fluctuations}

The average energy current alone does not fully characterize nonequilibrium transport; its statistics also depend on current fluctuations and higher-order cumulants. To quantify the steady-state transport precision, we include the time-normalized signal-to-noise ratio associated with reservoir $\mu$~\cite{Guarnieri2019prr,Engelhardt2024,Jacob2025}
\begin{eqnarray}\label{eq19}
\mathcal S_\mu=\frac{|J_\mu|}{\sqrt{J^{(2)}_\mu}},
\end{eqnarray}
where \(J\) and \(J^{(2)}\) denote the first and second current cumulants per unit time, respectively
Using the driven quantum master equation with FCS, we examine the second-order current fluctuations $J^{(2)}_{r}$, the time-normalized signal-to-noise ratio $\mathcal{S}_{r}$, and the third cumulant $J^{(3)}_{r}$ to further characterize transport.
Single-photon driving ($N_d=1$) is used as a reference to assess how two-photon driving ($N_d=2$) modifies higher-order transport statistics.

Fig.~\ref{fig:fig4} shows the dependence of the second-order current fluctuations, time-normalized signal-to-noise ratio, and third current cumulant on the Kerr nonlinearity $U$ and driving frequency $\omega_d$.
Since the current fluctuations at the drive input cannot be counted directly in this approach, we instead analyze fluctuations of the energy current into the right reservoir.
As shown in Fig.~\ref{fig:fig4}(a), $J^{(2)}_{\rm{r}}$ exhibits distinct stripe-like resonance structures in parameter space.
These resonance branches have a distribution similar to the regions of enhanced drive input current discussed above. Thus, when the driving frequency meets the resonance condition set by the Kerr energy-level spacings, system-reservoir energy exchange enhances both the average current and its fluctuations.
We also compare the second-order fluctuations for different driving photon orders.
Fig.~\ref{fig:fig4}(b) shows that the peaks in $J^{(2)}_{\rm{r}}$ near resonance positions are more pronounced under two-photon driving than under single-photon driving, indicating stronger energy-exchange fluctuations, based on the efficient hybridization of Fock states, e.g., at Eqs.~(\ref{p2r}-\ref{p0r}).
Figs.~\ref{fig:fig4}(c) and \ref{fig:fig4}(d) show that $\mathcal{S}_{\rm{r}}$ also increases in near-resonance regions. Although both the average current and the second-order fluctuations are enhanced near resonance, the larger relative increase in the average current yields a higher signal-to-noise ratio and improves the statistical distinguishability of the transport signal.

Furthermore, Figs.~\ref{fig:fig4}(e) and \ref{fig:fig4}(f) show the response of the third current cumulant $J^{(3)}_{\rm{r}}$ to driving.
The third cumulant characterizes departures of the current-fluctuation distribution from a Gaussian form and reveals higher-order fluctuations in nonequilibrium energy exchange.
Under two-photon driving, $J^{(3)}_{\rm{r}}$ also exhibits enhancement closely associated with resonances in the nonlinear energy-level structure, with more pronounced peaks than under single-photon driving.
Thus, two-photon driving modifies not only the average current and second-order fluctuations but also the non-Gaussian features of the energy-exchange distribution.
Fig.~\ref{fig:fig4} therefore demonstrates that Kerr nonlinearity and two-photon driving jointly control current fluctuations, the signal-to-noise ratio, and higher-order current cumulants near resonance, providing a richer set of fluctuation signatures for identifying nonequilibrium transport under two-photon driving.

\begin{figure*}[t]
    \centering
    \includegraphics[width=\linewidth]{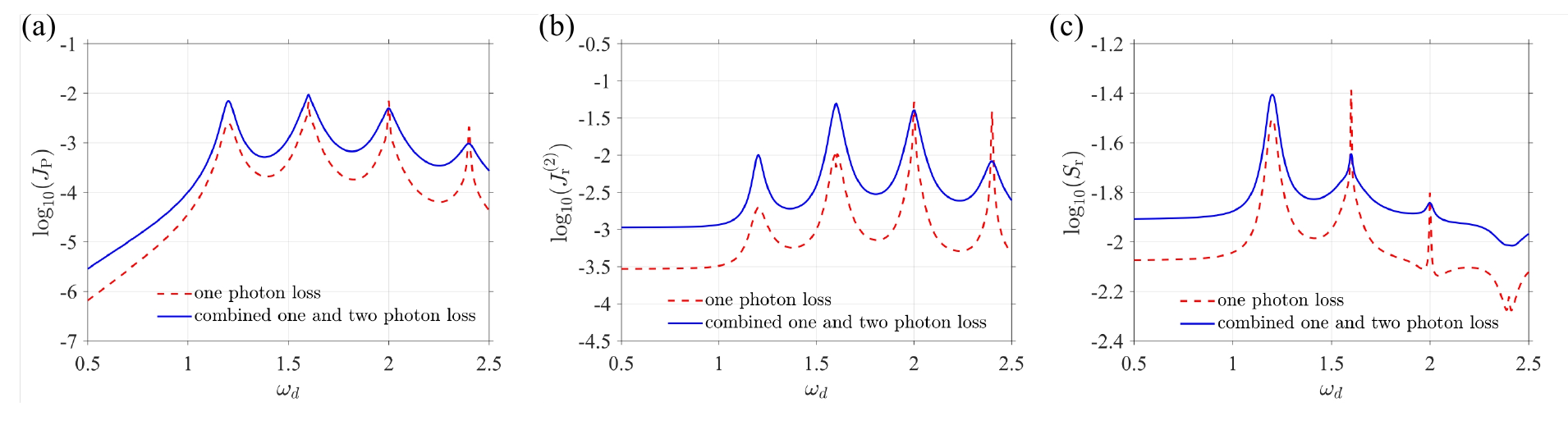}
\caption{Effect of two-photon loss on nonequilibrium transport in the Kerr resonator. (a) Drive input energy current $\log_{10}(J_{\rm{P}})$, (b) second-order current fluctuations $\log_{10}(J^{(2)}_{\rm{r}})$, and (c) signal-to-noise ratio $\log_{10}(\mathcal{S}_r)$ as functions of driving frequency $\omega_d$. The red dashed line represents single-photon loss alone, and the blue solid line represents coexisting single- and two-photon loss channels. The parameters are $N_{\rm{ph}}=30$, $\omega_0=1$, $\eta=0.02\omega_0$, $U=0.2\omega_0$, $\alpha_{l}=\alpha_{r}=0.001$ for both single- and two-photon dissipation, $\omega_c=10\omega_0$, $k_{\rm{B}}T_l=0.8\omega_0$, and $k_{\rm{B}}T_r=0.4\omega_0$.}
    \label{fig:fig5}
\end{figure*}

\subsection{Two-photon loss}
Two-photon loss has recently attracted considerable attention in quantum light--matter systems~\cite{Malekakhlagh2019prl,Li2024prl,Shah2025prl,Xiong2025prr}, where it produces behavior markedly different from conventional single-photon loss.
For independent single- and two-photon couplings to the reservoirs, the dissipative interaction is
\begin{equation}
\hat{V}_\mu=\sum_{k,\nu=1,2}[g_{k,\mu,\nu}\hat{b}^\dag_{k\mu}\hat{a}^\nu+g^{*}_{k,\mu,\nu}(\hat{a}^{\dag})^\nu\hat{b}_{k\mu}].\label{vRp2}
\end{equation}
After the rotating-frame transformation, the photon--reservoir interaction in the interaction picture becomes
$\hat{V}^\textrm{R}_{\mu,I}(t)=\sum_{k,\nu}[g_{k,\mu,\nu}e^{i(\omega_{k,\mu}-\nu\omega_d)t}\hat{b}^\dag_{k\mu}\hat{a}^\nu_{I}(t)+\rm{H.c.}]$.
The corresponding driven quantum master equation is derived in Appendix~\ref{app:loss},
which shows that incoherent transitions associated with two-photon loss involve $\gamma_{\mu}(2\omega_{d}+E_{mn})$ and $n_{\mu}(2\omega_{d}+E_{mn})$.
Compared with single-photon loss, two-photon loss increases the number of photons involved in system-reservoir energy exchange.
This directly affects the frequency dependence of the steady-state energy current, its fluctuations, and the signal-to-noise ratio. In the numerical comparisons below, the two-photon loss channels are added to the existing single-photon loss channels; single-photon loss remains present throughout.

Fig.~\ref{fig:fig5} compares nonequilibrium transport in the Kerr resonator with single-photon loss alone and with additional two-photon loss.
In Fig.~\ref{fig:fig5}(a), the drive input energy current $J_{\rm P}$ exhibits drive-induced enhancement for both loss configurations.
Adding two-photon loss, however, produces a stronger current response and more pronounced peaks near resonance.
This indicates that the external two-photon drive and the two-photon dissipation channels open additional energy-exchange pathways and produce a larger current response near resonance.
Fig.~\ref{fig:fig5}(b) shows that adding two-photon loss also amplifies the current fluctuations $J^{(2)}_{\rm{r}}$, producing particularly pronounced peaks near resonance.
This additional nonlinear loss channel does not merely rescale the single-photon dissipation rate; it modifies the frequency-dependent transition structure and hence the current fluctuations.
Fig.~\ref{fig:fig5}(c) further shows that the time-normalized signal-to-noise ratio ($\mathcal{S}_{\rm{r}}$) is generally higher when both single- and two-photon loss are present than with single-photon loss alone.
We conclude that within the parameter regime considered here, adding two-photon loss enhances the average current and increases the signal-to-noise ratio near several resonances.
It is consequently of considerable interest for studies of transport and fluctuations in quantum optical systems.

\begin{figure*}[t]
    \centering
    \includegraphics[width=\linewidth]{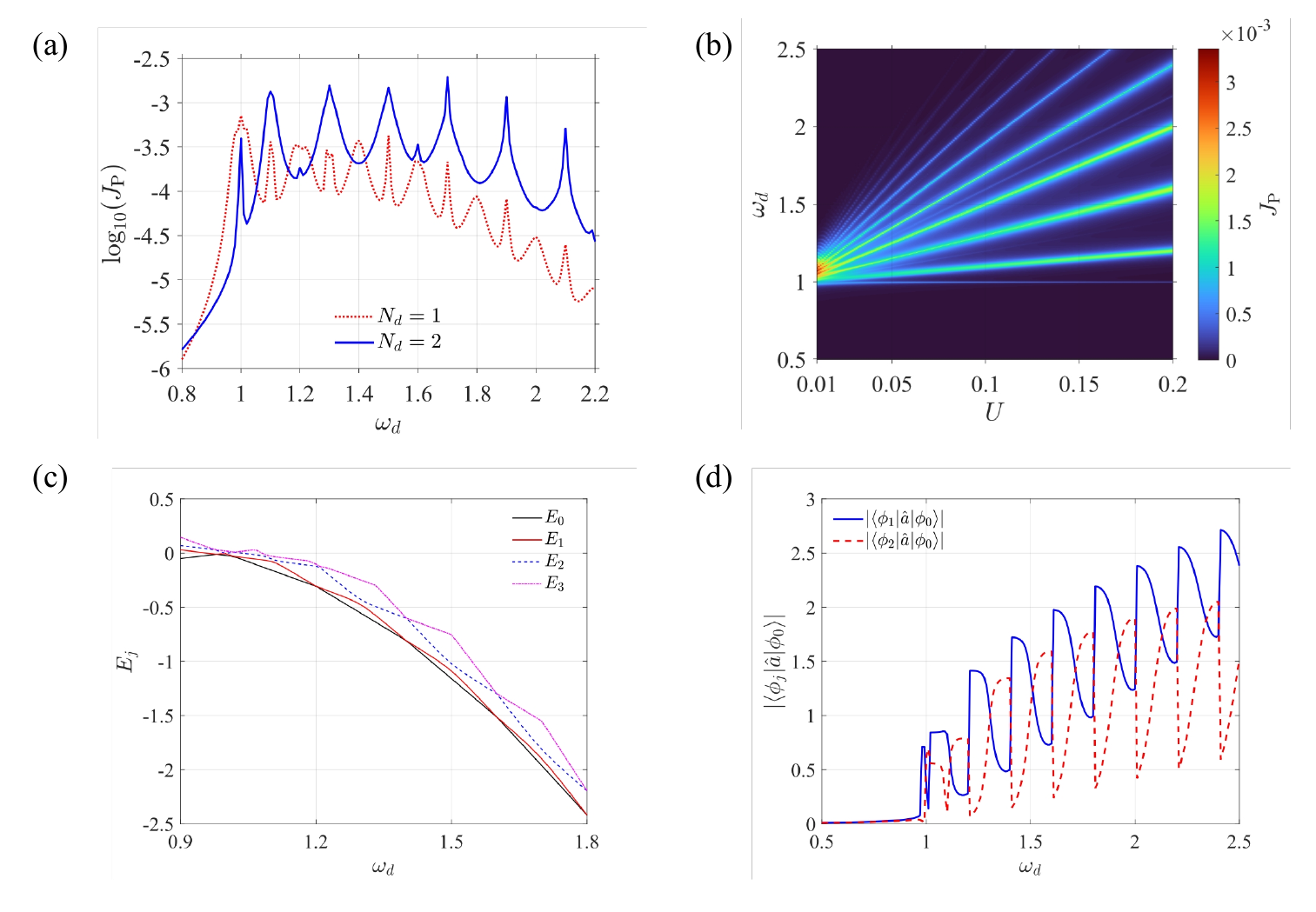}
\caption{Drive input energy current in the two-photon-driven nonlinear Jaynes--Cummings model. (a) Comparison of the input energy current as a function of driving frequency under two-photon and single-photon driving. (b) Drive input energy current $J_{\rm{P}}$ as a function of Kerr nonlinearity $U$ and driving frequency $\omega_d$. (c) Low-lying energy levels as functions of $\omega_d$. (d) Transition coupling coefficients $|{\langle}\phi_j|\hat{a}|\phi_0{\rangle}|$ as functions of $\omega_d$. For panels (a), (c), and (d), $U=0.1\omega_0$. The remaining parameters are $N_{\rm{ph}}=30$, $\omega_0=1$, qubit splitting $\varepsilon=\omega_0$, coupling strength $\lambda=0.02\omega_0$, $\eta=0.01\omega_0$,  $\alpha_{l}=\alpha_{r}=0.001$, $\omega_c=10\omega_0$, $k_{\rm{B}}T_l=1.2\omega_0$, and $k_{\rm{B}}T_r=0.6\omega_0$.}
    \label{fig:fig6}
\end{figure*}

\section{Two-photon-driven nonlinear Jaynes-Cummings model}\label{sec:jcm}
We now introduce Jaynes-Cummings coupling into the Kerr resonator~\cite{JCM1963pieee,Braak2016jpa} and investigate transport fluctuations under two-photon driving.
For the nonlinear Jaynes-Cummings model, we consider two independent dissipation channels: the cavity field couples to the left reservoir through the photon operators $\hat{a}^\dag~(\hat{a})$, while the two-level system couples to the right reservoir through the raising and lowering operators $\hat{\sigma}_+~(\hat{\sigma}_-)$.
The system transition operators in Eq.~(\ref{vR0}) are therefore $\hat A_l=\hat a$ and $\hat A_r=\hat\sigma_-$, respectively.
With two-photon driving, the rotating-frame system Hamiltonian is
\begin{align}\label{Hdjcm}
    \hat{H}_{\rm dJCM}&=\Delta_p\hat{a}^\dag\hat{a}+\frac{\Delta_q}{2}\hat{\sigma}_z+\lambda(\hat{a}^\dag\hat{\sigma}_-+\hat{a}\hat{\sigma}_+)\nonumber\\
    &+U\hat{a}^\dag\hat{a}^\dag\hat{a}\hat{a}-{\eta}[(\hat{a}^\dag)^2+\hat{a}^2],
\end{align}
The detunings are $\Delta_p=\omega_0-\omega_d$ and $\Delta_q=\varepsilon-\omega_d$.
The two-photon-driven nonlinear Jaynes-Cummings system is governed by four contributions: the cavity and qubit detunings, coherent light--matter coupling, Kerr nonlinearity, and two-photon driving.
The drive term $-\eta[(\hat{a}^{\dagger})^2+\hat{a}^{2}]$ creates or annihilates cavity photons in pairs.
The Jaynes-Cummings coupling $\lambda(\hat{a}^{\dagger}\hat{\sigma}_{-}+\hat{a}\hat{\sigma}_{+})$, together with the Kerr nonlinearity, hybridizes the photon and two-level-system states, producing a more complex dressed-state energy spectrum.
In the absence of driving ($\eta=0$), it is convenient to return to the laboratory frame, equivalently setting $\omega_d=0$ in Eq.~(\ref{Hdjcm}). The corresponding eigenenergies are
\begin{equation}~\label{Ejcm}
    \mathcal{E}^\prime_{m,\pm}=(m-1/2)\omega_0+U(m-1)^2{\pm}\frac{1}{2}\sqrt{\delta^2_m+4\lambda^2m},~m{\ge}1
\end{equation}
where $\delta_m=\omega_0-\varepsilon+2U(m-1)$ is the energy bias. The corresponding eigenstates are
\begin{subequations}
    \begin{align}
    |m,+{\rangle}=&\cos\theta_m|m,\downarrow{\rangle}+\sin\theta_m|m-1,\uparrow{\rangle},~\label{jcmvec+}\\
    |m,-{\rangle}=&-\sin\theta_m|m,\downarrow{\rangle}+\cos\theta_m|m-1,\uparrow{\rangle},\label{jcmvec-}
    \end{align}
\end{subequations}
The mixing angle is defined by $\tan(2\theta_m)=2\lambda\sqrt{m}/\delta_m$.
For $m=0$, the energy is $\mathcal{E}^\prime_{0,-}=-\varepsilon/2$, with the corresponding state $|0,-{\rangle}=|0,\downarrow{\rangle}$.
When $U=0$, these expressions reduce to the standard Jaynes-Cummings model (JCM) solutions~\cite{Braak2016jpa}.

\begin{figure*}[t]
    \centering
    \includegraphics[width=\linewidth]{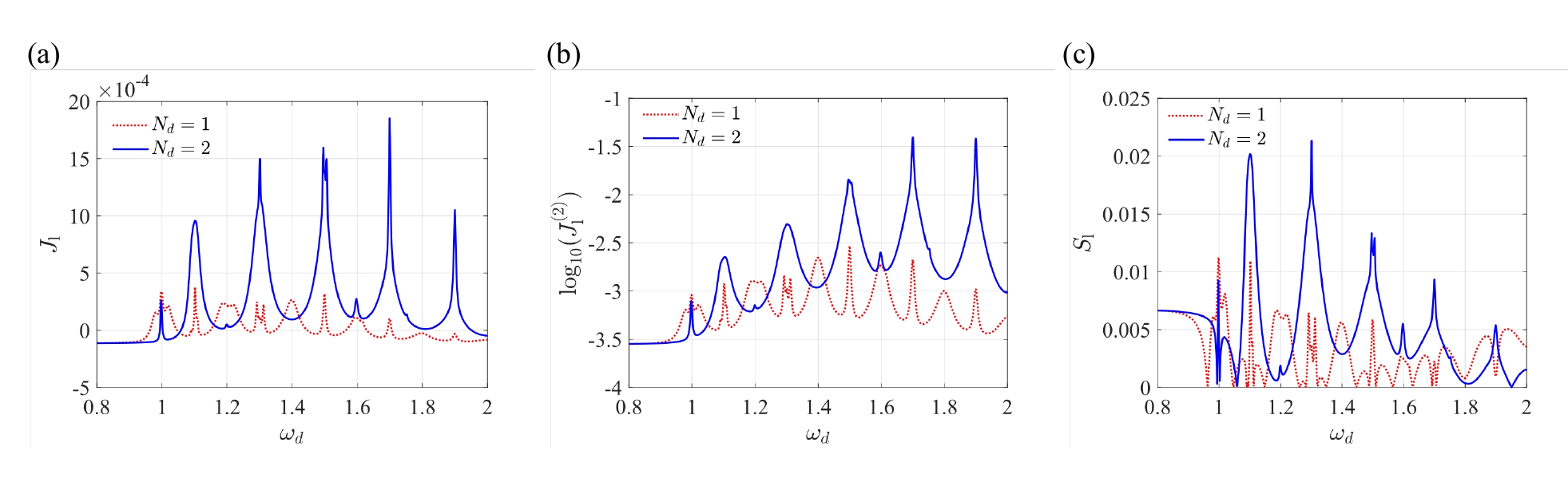}
\caption{Fluctuations of the energy current into the left reservoir in the photon-driven Jaynes--Cummings model. (a) Energy current $J_{l}$, (b) second-order current fluctuations $\log_{10}(J^{(2)}_{l})$, and (c) signal-to-noise ratio $\mathcal{S}_{l}$ as functions of driving frequency $\omega_d$. The red dashed and blue solid lines represent single-photon and two-photon driving, respectively. The other parameters are $N_{\rm{ph}}=30$, $\omega_0=1$, $U=0.1\omega_0$, qubit splitting $\varepsilon=\omega_0$, $\eta=0.01\omega_0$, coupling strength $\lambda=0.02\omega_0$, $\alpha_{l}=\alpha_{r}=0.001$, $\omega_c=10\omega_0$, $k_{\rm{B}}T_{l}=1.2\omega_0$, and $k_{\rm{B}}T_{r}=0.6\omega_0$.}
    \label{fig:fig7}
\end{figure*}

\subsection{Energy current under two-photon driving}

Fig.~\ref{fig:fig6} presents the drive input energy current in the two-photon-driven nonlinear Jaynes-Cummings model and its underlying mechanism.
We first compare steady-state current based on two-photon and single-photon drivings in Fig.~\ref{fig:fig6}(a),
which shows that the current peaks near several resonances are more pronounced under two-photon driving than single-photon case.
Thus, two-photon driving is more than a simple extension of single-photon driving.
Pairwise photon processes couple different dressed states and enable additional resonant transitions that are absent or weaker under single-photon driving.
Thus, two-photon driving can therefore induce a stronger nonequilibrium energy in the Jaynes-Cummings system.
Below, we focus on the nonequilibrium transport features under two-photon driving.

We next examine the dependence of $J_{\rm P}$ on driving frequency and Kerr nonlinearity, as shown in Fig.~\ref{fig:fig6}(b).
As in the Kerr resonator, $J_{\rm{P}}$ exhibits pronounced resonant enhancement. The undriven dressed states formed by Jaynes-Cummings coupling and Kerr nonlinearity, given in Eqs.~(\ref{jcmvec+}) and (\ref{jcmvec-}), produce a more complex energy-level structure.
In the weak-driving limit, the approximate condition for the $m$th group of resonances is
\begin{equation}
    2\omega_d{\approx}\mathcal{E}^\prime_{m+1,\sigma}-\mathcal{E}^\prime_{m-1,\sigma^\prime}.
\end{equation}
where $\sigma,\sigma'=\pm$ label the two Jaynes-Cummings dressed-state branches.
It is interesting to find that the resonance relation nonlinearly relies on the photon and qubit coupling structure, embedded in energies at Eq.~(\ref{Ejcm}), which is apparently distinct from the Kerr resonator counterpart.
Consequently, based on the spin-induced energy-level hybridization, the resonance branches in the energy current exhibit a richer peak structure as $\omega_d$ varies than in the Kerr resonator.

To clarify the origin of the current peaks, we consider low temperatures and the vicinity of the $m$th avoided crossing between the first and second excited states, as shown in Fig.~\ref{fig:fig6}(c).
Degenerate perturbation theory gives the approximate excited states
$|\phi_{2,m}{\rangle}{\approx}(|m+1,-{\rangle}-|m-1,-{\rangle})/\sqrt{2}$ and
$|\phi_{1,m}{\rangle}{\approx}(|m+1,-{\rangle}+|m-1,-{\rangle})/\sqrt{2}$,
while the ground state is $|\phi_{0,m}{\rangle}{\approx}|m,-{\rangle}$.
The corresponding eigenenergies are
${E}^\prime_{2,m}{\approx}(m+1/2)\Delta_p+Um^2{-}\frac{1}{2}\sqrt{\delta^2_{m+1}+4\lambda^2(m+1)}+\zeta_m$,
${E}^\prime_{1,m}{\approx}(m-3/2)\Delta_p+U(m-2)^2{-}\frac{1}{2}\sqrt{\delta^2_{m-1}+4\lambda^2(m-1)}-\zeta_m$,
with hybrid coefficient
$\zeta_m=\eta\sqrt{(m+1)}(\sqrt{m}\cos\theta_m\cos\theta_{m+2}+\sqrt{m+2}\sin\theta_m\sin\theta_{m+2})$
and ${E}^\prime_{0,m}{\approx}(m-1/2)\Delta_p+U(m-1)^2{-}\frac{1}{2}\sqrt{\delta^2_{m}+4\lambda^2m}$.
The approximate drive input energy current is  expressed as
$J_{\rm P}\approx
\sum_{j=1,2}\{(\omega_d+\Lambda_j)
[
\Gamma_-(\omega_d+\Lambda_j)P_j
-\Gamma_+(\omega_d+\Lambda_j)P_0
]+ (\omega_d-\Lambda_j)
[
\Gamma_-(\omega_d-\Lambda_j)P_0
-\Gamma_+(\omega_d-\Lambda_j)P_j
]\}$
,
where $\Lambda_j={E}^\prime_{j,m}-{E}^\prime_{0,m}$ and $P_j$ is the population of $|\phi_{j,m}{\rangle}$.
It is intriguing to find that under low excitation approximation, such expression of current is analogous to the Kerr-resonator case in Eq.~(\ref{Jp3TL}).
As expected,
such expression shows that the hybridization of the low-lying eigenstates near the resonance positions renders the relevant efficient incoherent transitions, thereby activating additional energy exchange channels.
In Fig.~\ref{fig:fig6}(d), it is found that ${\langle}\phi_{j}|\hat{a}|\phi_{0}{\rangle}{\neq}0$, where the resonance index $m$ has been omitted.
At the same time, $P_{1}$ and $P_{2}$ remain appreciable near resonance, leading to a substantial enhancement of the input energy current under two-photon driving.
We therefore conclude that this enhancement in the nonlinear Jaynes-Cummings model arises primarily from resonant coupling between the drive and the dressed-state energy-level structure and is closely associated with drive-induced population transfer among the dressed states.

\begin{figure*}[t]
    \centering
    \includegraphics[width=\linewidth]{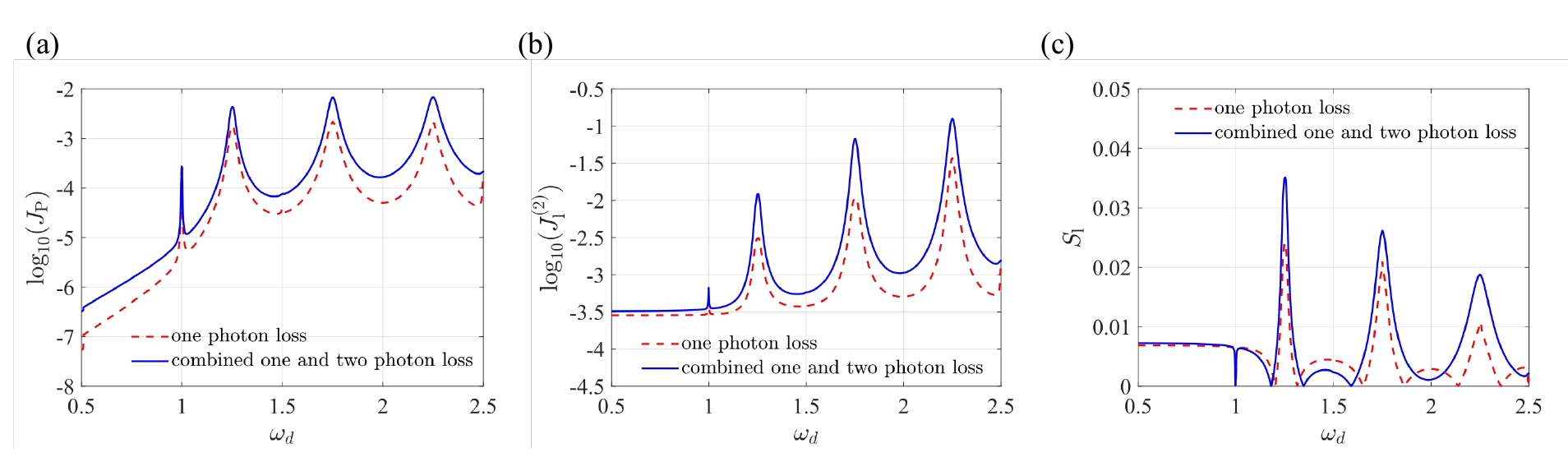}
\caption{Effect of two-photon loss on nonequilibrium transport in the driven nonlinear Jaynes-Cummings model. (a) Drive input energy current $\log_{10}(J_{\rm{P}})$, (b) second-order current fluctuations $\log_{10}(J^{(2)}_{l})$, and (c) signal-to-noise ratio $\mathcal{S}_l$ as functions of driving frequency $\omega_d$. The red dashed line represents single-photon dissipation alone, and the blue solid line represents coexisting single- and two-photon dissipation channels. The parameters are $N_{\rm{ph}}=30$, photon frequency $\omega_0=1$, qubit splitting $\varepsilon=\omega_0$, $\eta=0.01\omega_0$, $U=0.25\omega_0$, coupling strength $\lambda=0.02\omega_0$, $\alpha_{l}=\alpha_{r}=0.001$ for both single- and two-photon dissipation, $\omega_c=10\omega_0$, $k_{\rm{B}}T_{l}=1.2\omega_0$, and $k_{\rm{B}}T_{r}=0.6\omega_0$.}
    \label{fig:fig8}
\end{figure*}

\subsection{Time-normalized signal-to-noise ratio}

According to our sign convention, $J_l<0$ denotes energy flowing from the left reservoir into the system. In Fig.~\ref{fig:fig7}(a), we plot the current  $J_l$ on a logarithmic scale.
Since current fluctuations cannot be described directly at the drive input in this approach, we instead analyze fluctuations of the energy current into the left reservoir.
Fig.~\ref{fig:fig7} shows the energy current $J_l$, second current cumulant $J^{(2)}_l$, and signal-to-noise ratio ${\mathcal{S}}_l$ as functions of driving frequency $\omega_d$ in the two-photon-driven Jaynes-Cummings model.
All three quantities exhibit clear resonance structures, with more pronounced responses under two-photon driving than under single-photon driving.
Compared with the Kerr resonator, light-matter coupling introduces additional energy-level structure and more closely spaced resonance channels, resulting in a richer multipeak response.
In Fig.~\ref{fig:fig7}(a), the magnitude $J_{l}$ is strongly enhanced near several resonances. When the drive approaches a dressed-state resonance, for example,
one involving the transitions
$|\phi_0\rangle\leftrightarrow|\phi_1\rangle$
or
$|\phi_0\rangle\leftrightarrow|\phi_2\rangle$, two-photon driving produces sharper current peaks than single-photon driving.
This behavior closely resembles that of the drive input current $J_{\rm P}$, showing that the current enhancement induced by two-photon driving can also be observed at the reservoirs.
The second-order current fluctuations $J^{(2)}_l$ in Fig.~\ref{fig:fig7}(b) are similarly enhanced under two-photon driving, as is $J_l$, indicating a simultaneous increase in microscopic system-reservoir energy exchange and its fluctuations.
Fig.~\ref{fig:fig7}(c) further shows that the signal-to-noise ratio $\mathcal{S}_{l}$ is higher near several resonances under two-photon driving, with the enhancement occurring at frequencies close to the fluctuation peaks.
Although the increases in average current and fluctuations compete, 
the relative enhancement of the mean current exceeds that of the fluctuation noise in these resonance regions, resulting in a higher signal-to-noise ratio.
Similar results are obtained for the energy current into the right reservoir.
Thus, two-photon driving can enhance the average current and improve the signal-to-noise ratio simultaneously across several resonance channels, with an overall response stronger than that of single-photon driving.

\subsection{Two-photon loss}

We extend the preceding single-photon dissipation model by adding two-photon dissipation channels, as in Eq.~(\ref{vRp2}), to investigate how their combined action affects nonequilibrium transport in the nonlinear Jaynes-Cummings model.
In Fig.~\ref{fig:fig8}(a), $J_{\rm{P}}$ exhibits a pronounced multipeak resonance structure both with single-photon dissipation alone and with coexisting single- and two-photon dissipation.
Adding two-photon loss produces a stronger current signal over a broad range of driving frequencies.
This indicates that the additional photon-pair dissipation channels provide new system--reservoir energy-exchange pathways under two-photon driving, thereby enhancing the drive input energy current. The same trend is observed for $J_l$.
Fig.~\ref{fig:fig8}(b) shows that adding two-photon dissipation increases $J^{(2)}_{l}$ substantially near several resonances compared with single-photon loss alone. Two-photon dissipation therefore amplifies energy-exchange fluctuations while enhancing the average current.
Rather than simply increasing the dissipation strength, two-photon loss introduces additional transition matrix elements associated with the operator $\hat a^2$ and samples the reservoir spectrum at the shifted frequencies $2\omega_d+E_{mn}$, thereby redistributing the weights of energy-exchange processes.
Fig.~\ref{fig:fig8}(c) shows that once two-photon dissipation is added, the signal-to-noise ratio $\mathcal{S}_l$ is generally higher over a wide driving-frequency range than with single-photon dissipation alone.
This suggests that two-photon loss has a pronounced influence on the transport statistics under two-photon driving and can increase the signal-to-noise ratio over several frequency intervals.
We conclude that adding two-photon dissipation to the existing single-photon dissipation channels enhances the current, its fluctuations, and the signal-to-noise ratio over a broad frequency range, with all three displaying stronger responses than in the single-photon dissipation case.
The photon order of the dissipation channels is therefore another important factor in controlling quantum transport fluctuations in light-matter coupled systems.

\section{Possible experimental realization}~\label{sec:realization}
One proof-of-principle realization could employ a two-port
superconducting SNAIL resonator, 
which combines tunable Kerr
nonlinearity with a parametric two-photon
drive~\cite{He2023,Frattini2018} and follows superconducting-circuit
proposals for driven Kerr resonators~\cite{Puri2017}.
Two microwave environments with independently controlled
effective temperatures would establish the thermal
bias~\cite{Ronzani2018np}.
Illustrative operating regimes are
$\omega_0/(2\pi)\simeq4.5\,\mathrm{GHz}$,
$U/(2\pi)\simeq1\,\mathrm{MHz}$,
$\eta/(2\pi)\simeq0.1\,\mathrm{MHz}$, and
$\kappa_{l,r}/(2\pi)\sim0.1\,\mathrm{MHz}$, with
$T_l\simeq170\,\mathrm{mK}$ and $T_r\simeq20\,\mathrm{mK}$.
Sweeping the pump frequency $\omega_p=2\omega_d$ would search
for resonances near $\omega_d=\omega_0+(2m-1)U
~(m=1,2,\ldots)$.
Their visibility is determined by the
calibrated drive and dissipation.

Transport fluctuations could be probed through time-resolved
energy-transfer trajectories at one terminal~\cite{Yang2020nc}.
Repeated photon detections would generate time-binned
photon-counting records~\cite{Lescanne2020}, from which the accumulated
transferred energy may be written as
$Q_r(t)\simeq\hbar\omega_0 N_r(t)$,
with negligible target-terminal thermal occupation and narrowband emission.
Repeated steady-state trajectories then could effectively determine the distribution
$P_t(Q_r)$, whose long-time cumulants yield the mean current and current
fluctuations~\cite{Wang2021pra}, e.g., for the right terminal
$J_r=\langle Q_r\rangle/t$,
$J_r^{(2)}=\langle(\delta Q_r)^2\rangle/t$, and
$J_r^{(3)}=\langle(\delta Q_r)^3\rangle/t$.
Scanning the two-photon pump frequency would therefore allow the predicted
resonance dependence of both the average energy current and its fluctuations
to be tested. 

\section{Conclusions}\label{sec:conclusions}

Formulating a driven quantum master equation in the rotating frame combined with full counting statistics, we have investigated nonequilibrium energy transport and its fluctuations in representative nonequilibrium quantum optical models under two-photon driving.
By retaining the drive-induced frequency shifts in the system--reservoir interaction, this framework provides access to the drive input energy current, second-order current fluctuations, and third current cumulant.

For the two-photon-driven Kerr resonator, the driven quantum master equation agrees closely with the Floquet-master-equation results for steady-state transport.
The interplay between Kerr-induced anharmonicity and two-photon driving produces a pronounced multipeak resonance structure in the input energy current. 
In the weak-driving regime, the $m$th group of two-photon resonances satisfies $2\omega_d\simeq E_{m+1}-E_{m-1}$, yielding the approximate resonance positions $\omega_d^{(m)}\simeq\omega_0+(2m-1)U$.
At low temperatures, retaining the three lowest dressed states, able to capture resonance feature,
further yields an approximate analytical expression for the drive input energy current. 
The dominant contributions to $J_P$ arise from two sets of drive-assisted incoherent transition channels involving the energies $\omega_d+\Lambda_j$ and $\omega_d-\Lambda_j$.
Near resonance, two-photon-drive-induced hybridization of the low-energy dressed states renders the relevant transition matrix elements nonzero.
Together with the redistribution of steady-state populations, this activates the incoherent transition channels and enhances the drive input current.
Compared with single-photon driving, two-photon driving enhances the average energy current and substantially modifies the fluctuations and non-Gaussian features of energy exchange. 
Near certain resonances, tuning the driving frequency can enhance both the average current and time-normalized signal-to-noise ratio.
Adding two-photon dissipation to the existing single-photon dissipation channels introduces additional photon-pair exchange pathways and modifies the associated energy-exchange statistics, enhancing the drive input current near several resonances.
Two-photon dissipation also substantially modifies second-order current fluctuations and improves time-normalized signal-to-noise ratio.

In the nonlinear Jaynes-Cummings model, light-matter couplings produce an additional dressed-state structure.
The two-photon resonance condition generalizes to $2\omega_d\simeq E'_{m+1,\sigma}-E'_{m-1,\sigma'}$, which nonlinearly depends on photon and qubit coupling structure.
Thus, it produces a richer set of resonance branches than in the Kerr resonator.
{In analogy with the Kerr resonator case, a low-energy-state approximation further also provides an effective description of the drive input current near resonance, while two-photon dissipation also modifies the current fluctuations and time-normalized signal-to-noise ratio.}
The present framework can, in principle, be extended to transport fluctuations in other driven-dissipative platforms, including optomechanical systems.

These results clarify how photon driving, nonlinear dressed-state structure, and incoherent energy exchange processes jointly shape nonequilibrium energy transport and its fluctuations, and provide a basis for exploring transport control through multiphoton processes in quantum optical systems.

\begin{acknowledgments}
This work was supported by the National Natural Science Foundation of China (Grant No. 12674313 and No. 12305050), the Zhejiang Provincial Natural Science Foundation of China (Grant No. LZ25A050001),
and the Natural Science Foundation of Jiangsu Higher Education Institutions of China (Grant No. 23KJB140017).
\end{acknowledgments}

\vspace{1.0cm}
\textbf{Data availability statement}

The numerical simulation codes used in this study are available at~\cite{Chen2026}.

\appendix
\section{Approximate energy-current expressions under two-photon driving}\label{app:currents}

For sufficiently weak system-reservoir coupling and within the parameter regime considered here, the steady-state coherences in the eigenbasis of $\hat{H}_{\rm{S},\rm{R}}$, without the effect of bath-induce coherence~\cite{Li2015aop}, may be approximately negligible compared with the populations.
The driven quantum master equation~(\ref{qme1}) then effectively reduces to
\begin{subequations}
\begin{align}
\frac{d}{dt}\hat{\rho}_{s,\rm{R}}(t)=&
-i[\hat{H}_{\rm{S},\rm{R}},\hat{\rho}_{s,\rm{R}}(t)]+\sum_\mu\mathcal{L}_\mu[\hat{\rho}_{s,\rm{R}}(t)],~\label{dDME}\\
\mathcal{L}_\mu[\hat{\rho}_{s,\rm{R}}(t)]=&\sum_{m,m^\prime}
\{\Gamma^\mu_{-}(\omega_d+E_{m^{\prime}m})\mathcal{D}[|\phi_m{\rangle}{\langle}\phi_{m^\prime}|]\hat{\rho}_{s,\rm{R}}(t)\nonumber\\
&+\Gamma^\mu_{+}(\omega_d+E_{m^{\prime}m})\mathcal{D}[|\phi_{m^\prime}{\rangle}{\langle}\phi_{m}|]\hat{\rho}_{s,\rm{R}}(t)\},    ~\label{dDMErate}    
    \end{align}
\end{subequations}
where the dissipator is defined as
\begin{equation}
    \mathcal{D}[\hat{O}]\hat{\rho}_{s,\rm R}=(2\hat{O}\hat{\rho}_{s, \rm R}\hat{O}^\dag-\hat{O}^\dag\hat{O}\hat{\rho}_{s,\rm R}-\hat{\rho}_{s,\rm R}\hat{O}^\dag\hat{O})/2.
\end{equation}
The eigenstates satisfy $\hat{H}_{\rm S,\rm R}|\phi_m{\rangle}=E_m|\phi_m{\rangle}$, and the energy differences are $E_{m^\prime{m}}=E_{m^\prime}-E_m$.
The transition rates are
\begin{subequations}
	\begin{align}
\Gamma^\mu_{-}(\omega_d+E_{m^{\prime}m})&=\gamma_\mu(\omega_d+E_{m^{\prime}m})[1+n_\mu(\omega_d+E_{m^{\prime}m})]\nonumber\\*&\quad\times |{\langle}\phi_m|\hat{A}_\mu|\phi_{m^\prime}{\rangle}|^2,~\label{R-}\\
\Gamma^\mu_{+}(\omega_d+E_{m^{\prime}m})&=\gamma_\mu(\omega_d+E_{m^{\prime}m})n_\mu(\omega_d+E_{m^{\prime}m})\nonumber\\*&\quad\times |{\langle}\phi_m|\hat{A}_\mu|\phi_{m^\prime}{\rangle}|^2.~\label{R+}
\end{align}
\end{subequations}
Here $n_\mu(\omega)=1/[\exp(\omega/k_{\textrm{B}}T_\mu)-1]$ is the Bose--Einstein distribution, and we use the Ohmic spectral function $\gamma_\mu(\omega)=\pi\alpha_\mu\theta(\omega)\omega\exp(-\omega/\omega_c)$, where $\alpha_\mu$ is the dissipation strength, $\omega_c$ is the reservoir cutoff frequency, and $\theta(\omega)$ is the Heaviside step function, with $\theta(\omega>0)=1$ and $\theta(\omega{\leq}0)=0$.
The incoherent rates $\Gamma^\mu_{-}(\omega_d+E_{m^{\prime}m})~(\Gamma^\mu_{+}(\omega_d+E_{m^{\prime}m}))$ describe downward and upward transitions from $|\phi_{m^\prime}{\rangle}~(|\phi_{m}{\rangle})$ to $|\phi_{m}{\rangle}~(|\phi_{m^\prime}{\rangle})$, respectively, with the reservoir exchanging an energy $\omega_d+E_{m^{\prime}m}$.
The corresponding steady-state energy current into reservoir $\mu$ is
\begin{eqnarray}~\label{JmuApp}
    J_\mu&=&\sum_{m,m^\prime}(\omega_d+E_{m^\prime{m}})[\Gamma^\mu_{-}(\omega_d+E_{m^\prime{m}})P_{m^\prime}\nonumber\\
    &&\qquad-\Gamma^\mu_{+}(\omega_d+E_{m^\prime{m}})P_{m}],
\end{eqnarray}
where $P_m$ is the steady-state population of the eigenstate $|\phi_m{\rangle}$.
Energy conservation gives the current at the driving terminal as
\begin{equation}~\label{JpApp}
    J_{\rm P}=J_l+J_r.
\end{equation}

At low temperatures, we retain only the three lowest steady-state populations, $P_0$, $P_1$, and $P_2$, to obtain the following low-energy approximations to  Eqs.~(\ref{JmuApp}) and (\ref{JpApp}):

\begin{widetext}
\begin{subequations}
\begin{align}
J_l&\approx\sum_{j=1}^{2} \Big\{ (\omega_d+\Lambda_j) [\Gamma^l_-(\omega_d+\Lambda_j)P_j -\Gamma^l_+(\omega_d+\Lambda_j)P_0] +(\omega_d-\Lambda_j) [\Gamma^l_-(\omega_d-\Lambda_j)P_0 -\Gamma^l_+(\omega_d-\Lambda_j)P_j] \Big\}\nonumber\\*& +(\omega_d+\Lambda_{21}) [\Gamma^l_-(\omega_d+\Lambda_{21})P_2 -\Gamma^l_+(\omega_d+\Lambda_{21})P_1] +(\omega_d-\Lambda_{21}) [\Gamma^l_-(\omega_d-\Lambda_{21})P_1 -\Gamma^l_+(\omega_d-\Lambda_{21})P_2] \nonumber\\*&+\omega_d\sum_{n=0}^{2} [\Gamma^l_{-,n}(\omega_d) -\Gamma^l_{+,n}(\omega_d)]P_n,\label{eqA6a}\\
J_r&\approx\sum_{j=1}^{2} \Big\{ (\omega_d+\Lambda_j) [\Gamma^r_-(\omega_d+\Lambda_j)P_j -\Gamma^r_+(\omega_d+\Lambda_j)P_0] +(\omega_d-\Lambda_j) [\Gamma^r_-(\omega_d-\Lambda_j)P_0 -\Gamma^r_+(\omega_d-\Lambda_j)P_j] \Big\}\nonumber\\*& +(\omega_d+\Lambda_{21}) [\Gamma^r_-(\omega_d+\Lambda_{21})P_2 -\Gamma^r_+(\omega_d+\Lambda_{21})P_1] +(\omega_d-\Lambda_{21}) [\Gamma^r_-(\omega_d-\Lambda_{21})P_1 -\Gamma^r_+(\omega_d-\Lambda_{21})P_2] \nonumber\\*&+\omega_d\sum_{n=0}^{2} [\Gamma^r_{-,n}(\omega_d) -\Gamma^r_{+,n}(\omega_d)]P_n,\label{eqA6b}\\
J_{\rm P}&\approx\sum_{j=1}^{2} \Big\{ (\omega_d+\Lambda_j) [\Gamma_-(\omega_d+\Lambda_j)P_j -\Gamma_+(\omega_d+\Lambda_j)P_0] +(\omega_d-\Lambda_j) [\Gamma_-(\omega_d-\Lambda_j)P_0 -\Gamma_+(\omega_d-\Lambda_j)P_j] \Big\}\nonumber\\*& +(\omega_d+\Lambda_{21}) [\Gamma_-(\omega_d+\Lambda_{21})P_2 -\Gamma_+(\omega_d+\Lambda_{21})P_1] +(\omega_d-\Lambda_{21}) [\Gamma_-(\omega_d-\Lambda_{21})P_1 -\Gamma_+(\omega_d-\Lambda_{21})P_2] \nonumber\\*&+\omega_d\sum_{n=0}^{2} [\Gamma_{-,n}(\omega_d) -\Gamma_{+,n}(\omega_d)]P_n.\label{eqA6c}
\end{align}
\end{subequations}
\end{widetext}
Here, the energy gaps are $\Lambda_j=E_j-E_0,~(j=1,2)$, and $\Lambda_{21}=E_2-E_1=\Lambda_2-\Lambda_1$.
The quantities $P_n$ are the steady-state populations of the three lowest eigenstates $|\phi_n\rangle$. 
The total rates are $\Gamma_{\pm}(\omega)=\sum_{\mu=l,r}\Gamma^\mu_{\pm}(\omega),$ and $\Gamma_{\pm,n}(\omega)=\sum_{\mu=l,r}\Gamma^\mu_{\pm,n}(\omega)$, with
$\Gamma^\mu_{\pm,n}(\omega_d)=\pm\gamma_\mu(\omega_d)n_\mu(\pm\omega_d)\left|\langle\phi_n|\hat A_\mu|\phi_n\rangle\right|^2$.
The rates involving $\omega_d\pm\Lambda_{21}$ in Eqs.~(\ref{eqA6a})--(\ref{eqA6c}) describe transitions between $|\phi_1\rangle$ and $|\phi_2\rangle$.

For the two-photon-driven Kerr resonator considered here, the Hamiltonian conserves photon-number parity, $[\hat H_{\rm Kerr},\hat\Pi]=0$, where $\hat\Pi=\exp(i\pi\hat a^\dagger\hat a)$.
The eigenstates $|\phi_n\rangle$ can therefore be chosen as eigenstates of $\hat\Pi$.
The system--reservoir coupling operator $\hat A_\mu=\hat a$ satisfies $\hat\Pi\hat a\hat\Pi^\dagger=-\hat a$, so its matrix elements between states of the same parity vanish.
Among the three lowest eigenstates considered near resonance, $|\phi_1\rangle$ and $|\phi_2\rangle$ have the same parity, whereas $|\phi_0\rangle$ has the opposite parity.
Consequently, $\langle\phi_n|\hat a|\phi_n\rangle=0,n=0,1,2$, and
$\langle\phi_1|\hat a|\phi_2\rangle=\langle\phi_2|\hat a|\phi_1\rangle=0$.
The definitions of the transition rates then give
$\Gamma_{\pm,n}^{\mu}(\omega_d)
=0~(n=0,1,2)$
and
$\Gamma_{\pm}^{\mu}(\omega_d+\Lambda_{21})=
\Gamma_{\pm}^{\mu}(\omega_d-\Lambda_{21})
=0$.

The low-energy approximation to the drive input energy current therefore reduces to
\begin{eqnarray}
J_{\rm P}\approx&
\sum_{j=1}^{2}
\Big\{
(\omega_d+\Lambda_j)
\left[
\Gamma_-(\omega_d+\Lambda_j)P_j
-\Gamma_+(\omega_d+\Lambda_j)P_0
\right]
\nonumber\\
&+(\omega_d-\Lambda_j)
\left[
\Gamma_-(\omega_d-\Lambda_j)P_0
-\Gamma_+(\omega_d-\Lambda_j)P_j
\right]
\Big\}.\nonumber\\
\label{eqA8}
\end{eqnarray}

A similar low-energy-state approximation applies to the two-photon-driven Jaynes-Cummings model with Kerr nonlinearity.
At low temperatures and near the $m$th avoided crossing of the low-lying levels, the two-photon drive predominantly couples $|m+1,-\rangle$ and $|m-1,-\rangle$.
The three relevant low-energy states retained in this local approximation are therefore approximately
\begin{subequations}
\begin{align}
|\phi_{2,m}\rangle
&\simeq
\frac{1}{\sqrt{2}}
\left(
|m+1,-\rangle-|m-1,-\rangle
\right),\\
|\phi_{1,m}\rangle
&\simeq
\frac{1}{\sqrt{2}}
\left(
|m+1,-\rangle+|m-1,-\rangle
\right),\\
|\phi_{0,m}\rangle
&\simeq
|m,-\rangle.
\end{align}
\end{subequations}
For the two-photon-driven nonlinear Jaynes-Cummings Hamiltonian, the
total-excitation parity
$\hat{\Pi}=\exp(i\pi\hat{N})$, with
$\hat{N}=\hat{a}^{\dagger}\hat{a}
+\hat{\sigma}_{+}\hat{\sigma}_{-}$, is conserved,
i.e., $[\hat{H}_{\rm dJCM},\hat{\Pi}]=0$.
The system--reservoir coupling operators
$\hat{A}_{l}=\hat{a}$ and $\hat{A}_{r}=\hat{\sigma}_{-}$
both change the excitation parity. Since
$|\varphi_{1,m}\rangle$ and $|\varphi_{2,m}\rangle$
have the same parity, whereas $|\varphi_{0,m}\rangle$
has the opposite parity, the diagonal matrix elements and the matrix
elements between $|\varphi_{1,m}\rangle$ and
$|\varphi_{2,m}\rangle$ vanish.
We define the corresponding energy gaps as $\Lambda_j=E'_{j,m}-E'_{0,m}, j=1,2.$
Substituting these states into Eq.~(\ref{eqA6c}),
the parity selection rules
eliminate the diagonal contributions and the transitions between
$|\varphi_{1,m}\rangle$ and $|\varphi_{2,m}\rangle$.
The resulting local low-energy approximation to the drive input energy
current near the selected $(-,-)$ resonance branch is

\begin{eqnarray}
J_{\rm P}&\approx&
\sum_{j=1,2}
\Big\{
(\omega_d+\Lambda_j)
\left[
\Gamma_-(\omega_d+\Lambda_j)P_j
-\Gamma_+(\omega_d+\Lambda_j)P_0
\right]
\nonumber\\
&&+
(\omega_d-\Lambda_j)
\left[
\Gamma_-(\omega_d-\Lambda_j)P_0
-\Gamma_+(\omega_d-\Lambda_j)P_j
\right]
\Big\}.\nonumber\\
\end{eqnarray}
where $\Gamma_{\pm}(\omega)=\sum_{\mu=l,r}\Gamma^{\mu}_{\pm}(\omega)$, and the system transition operators associated with the left and right reservoirs are $\hat A_l=\hat a$ and $\hat A_r=\hat\sigma_-$, respectively.
Thus, near the selected $(-,-)$ resonance branch, the approximate drive
input current has the same formal structure as Eq.~(\ref{eqA8}), while the
transition rates are determined jointly by the photonic and
two-level-system dissipation channels and the Jaynes-Cummings
dressed-state structure. The other resonance branches associated with
different combinations of $(\sigma,\sigma')$ can be treated analogously
within their corresponding local low-energy subspaces.

\section{Driven quantum master equation with two-photon dissipation}\label{app:loss}
For the two-photon-driven Kerr resonator, we consider independent single- and two-photon dissipative couplings to the reservoirs.
The corresponding photon-reservoir interaction is

\begin{equation} 
\hat{V}_{\mu} = \sum_{k,\nu=1,2} \left(g_{k,\mu,\nu}\hat{b}^{\dagger}_{k\mu}\hat{A}_{\mu,\nu} + g^{*}_{k,\mu,\nu}\hat{A}^{\dagger}_{\mu,\nu}\hat{b}_{k\mu} \right), 
\end{equation} 
where $\hat{A}_{\mu,\nu}=\hat{a}^{\nu}$ for $\nu=1,2$. For simplicity, we assume that the single- and two-photon
dissipative channels have the same Ohmic spectral function,$\gamma_{\mu,1}(\omega)=\gamma_{\mu,2}(\omega)\equiv\gamma_{\mu}(\omega).$
The resulting driven master equation is

\begin{subequations}
\begin{align}~\label{qme2}
 	d\hat{\rho}_{s,\rm R}(t)/dt=&-i[\hat{H},\hat{\rho}_{s,\rm R}]
 	+\sum_{\mu=l,r;\nu=1,2}\mathcal{L}_{\mu,\nu}[\hat{\rho}_{s,\rm R}],\\
    \mathcal{L}_{\mu,\nu}[\hat{\rho}_{s,\rm R}] = \Big\{ &[\hat{D}^{\dagger}_{\mu,\nu,+}\hat{\rho}_{s,\rm R},\hat{A}_{\mu,\nu}] \nonumber\\ &+[\hat{D}_{\mu,\nu,-}\hat{\rho}_{s,\rm R},\hat{A}^{\dagger}_{\mu,\nu}] \Big\}+\mathrm{H.c.}.
\end{align}
\end{subequations}
Here, $\hat{\rho}_{s,\rm R}(t)=\hat{R}^\dag(t)\hat{\rho}_{s}(t)\hat{R}(t)$ is the system density operator in the rotating frame, and $\hat{A}_{\mu,\nu}=\hat{a}^{\nu}$ denotes the system operator associated with the $\nu$-photon dissipative channel.
The modified system operators are
\begin{subequations}
 	\begin{align}
\hat{D}_{\mu,\nu,+}&=\int^\infty_0d\tau\sum_k|g_{k,\mu,\nu}|^2n_{k,\mu}e^{-i(\omega_{k}-\nu\omega_d)\tau}\hat{A}_{\mu,\nu}(-\tau),~\label{Pmu+}\\
\hat{D}_{\mu,\nu,-}&=\int^\infty_0d\tau\sum_k|g_{k,\mu,\nu}|^2(1+n_{k,\mu})e^{-i(\omega_{k}-\nu\omega_d)\tau}\hat{A}_{\mu,\nu}(-\tau),~\label{Pmu-}
\end{align}
\end{subequations}
where $\hat{A}_{\mu,\nu}(-\tau)=e^{-i\hat{H}\tau}\hat{A}_{\mu,\nu}e^{i\hat{H}\tau}.$
Including FCS, the modified driven quantum master equation is given by
\begin{subequations}
\begin{align}~\label{DQMEFCS2}
 	d\hat{\rho}^\chi_{s,\rm R}(t)/dt=&-i[\hat{H},\hat{\rho}^\chi_{s,\rm R}]
 	+\sum_{\mu=l,r;\nu=1,2}\mathcal{L}_{\chi_\mu,\nu}[\hat{\rho}^\chi_{s,\rm R}],\\
\mathcal{L}_{\chi_\mu,\nu}[\hat{\rho}^\chi_{s,\rm R}]=&
[\hat{A}_{\mu,\nu}\hat{\rho}^\chi_{s,\rm R}\hat{D}^\dag_{\mu,\nu,-}(\chi_\mu)+\hat{A}^\dag_{\mu,\nu}\hat{\rho}^\chi_{s,\rm R}\hat{D}_{\mu,\nu,+}(\chi_\mu)\nonumber\\
&+\hat{D}_{\mu,\nu,-}(-\chi_\mu)\hat{\rho}^\chi_{s,\rm R}\hat{A}^\dag_{\mu,\nu}+\hat{D}^\dag_{\mu,\nu,+}(-\chi_\mu)\hat{\rho}^\chi_{s,\rm R}\hat{A}_{\mu,\nu}]\nonumber\\
&-[(\hat{A}_{\mu,\nu}\hat{D}^\dag_{\mu,\nu,+}\hat{\rho}^\chi_{s,\rm R}+\hat{A}^\dag_{\mu,\nu}\hat{D}_{\mu,\nu,-}\hat{\rho}^\chi_{s,\rm R})\nonumber\\
&+(\hat{\rho}^\chi_{s,\rm R}\hat{D}_{\mu,\nu,+}\hat{A}^\dag_{\mu,\nu}+\hat{\rho}^\chi_{s,\rm R}\hat{D}^\dag_{\mu,\nu,-}\hat{A}_{\mu,\nu})],
\end{align}
\end{subequations}
where the system dissipative operators are
\begin{subequations}
 	\begin{align}
\hat{D}_{\mu,\nu,+}(\chi_\mu)&=\sum_{n,m}\frac{\gamma_\mu(\nu\omega_d+E_{mn})}{2}n_\mu(\nu\omega_d+E_{mn})\nonumber\\*&\quad\times e^{-i(\nu\omega_d+E_{mn})\chi_\mu}A^{nm}_{\mu,\nu}|\varphi_n{\rangle}{\langle}\varphi_m|,~\label{Ppchi}\\
\hat{D}_{\mu,\nu,-}(\chi_\mu)&=\sum_{n,m}\frac{\gamma_\mu(\nu\omega_d+E_{mn})}{2}[1+n_\mu(\nu\omega_d+E_{mn})]\nonumber\\*&\quad\times e^{-i(\nu\omega_d+E_{mn})\chi_\mu}A^{nm}_{\mu,\nu}|\varphi_n{\rangle}{\langle}\varphi_m|~\label{Pmchi}.
\end{align}
\end{subequations}
where$A^{nm}_{\mu,\nu}=\langle\phi_n|\hat{A}^\nu_{\mu}|\phi_m\rangle=\langle\phi_n|\hat{a}^{\nu}|\phi_m\rangle$.

For the nonlinear Jaynes-Cummings model, both photonic and qubit dissipation channels are included. We use $p$ and $q$ to label the photonic and qubit reservoir channels, respectively.

\begin{subequations}
\begin{align}
\hat{V}_p=&\sum_k(g_{k,p,1}\hat{b}^\dag_{kp}\hat{a}+g^{*}_{k,p,1}\hat{b}_{kp}\hat{a}^{\dag})\nonumber\\
&+
\sum_k(g_{k,p,2}\hat{b}^\dag_{kp}\hat{a}^2+g^{*}_{k,p,2}\hat{b}_{kp}(\hat{a}^{\dag})^2),\label{App:vRp2}\\
\hat{V}_q=&\sum_k(g_{k,q}\hat{b}^\dag_{kq}\hat{\sigma}_-+g^{*}_{k,q}\hat{b}_{kq}\hat{\sigma}_+).
\end{align}
\end{subequations}
We define $\hat{A}_{p,\nu}=\hat{a}^{\nu}$ for the photonic channel and $\hat{A}_{q}=\hat{\sigma}_{-}$ for the qubit channel.
An analogous derivation yields the driven master equation
\begin{subequations}
\begin{align}~\label{qme2_jcm}
 	d\hat{\rho}_{s,\rm R}(t)/dt=&-i[\hat{H},\hat{\rho}_{s,\rm R}]
 	+\mathcal{L}_{q}[\hat{\rho}_{s,\rm R}]+\sum_{\nu=1,2}\mathcal{L}_{p,\nu}[\hat{\rho}_{s,\rm R}],\\
 \mathcal{L}_{q}[\hat{\rho}_{s,\rm R}]=&
 \{([\hat{D}^\dag_{q,+}\hat{\rho}_{s,\rm R},\hat{A}_q]
 	+[\hat{D}_{q,-}\hat{\rho}_{s,\rm R},\hat{A}^\dag_q])+\rm H.c.\},\\
    \mathcal{L}_{p,\nu}[\hat{\rho}_{s,\rm R}]=&
    \{([\hat{D}^\dag_{p,\nu,+}\hat{\rho}_{s,\rm R},\hat{A}_{p,\nu}]
 	+[\hat{D}_{p,\nu,-}\hat{\rho}_{s,\rm R},\hat{A}^{\dag}_{p,\nu}])\nonumber\\
    &+\rm H.c.\},
\end{align}
\end{subequations}
Here, $\hat{\rho}_{s,\rm R}(t)=\hat{R}^\dag(t)\hat{\rho}_{s}(t)\hat{R}(t)$ is the system density operator in the rotating frame, and the modified system operators are
\begin{subequations}
 	\begin{align}
\hat{D}_{q,+}&=\int^\infty_0d\tau\sum_k|g_{k,q}|^2n_{k,q}e^{-i(\omega_{k}-\omega_d)\tau}\hat{A}_q(-\tau),~\label{Qmu+}\\
\hat{D}_{q,-}&=\int^\infty_0d\tau\sum_k|g_{k,q}|^2(1+n_{k,q})e^{-i(\omega_{k}-\omega_d)\tau}\hat{A}_q(-\tau), ~\label{Qmu-}\\
\hat{D}_{p,\nu,+}&=\int^\infty_0d\tau\sum_k|g_{k,p,\nu}|^2n_{k,p}e^{-i(\omega_{k}-\nu\omega_d)\tau}\hat{A}_{p,\nu}(-\tau),~\label{Pmu+_jcm}\\
\hat{D}_{p,\nu,-}&=\int^\infty_0d\tau\sum_k|g_{k,p,\nu}|^2(1+n_{k,p})e^{-i(\omega_{k}-\nu\omega_d)\tau}\hat{A}_{p,\nu}(-\tau),~\label{Pmu-_jcm}
\end{align}
\end{subequations}
Consequently, the driven quantum master equation combined with full counting statistics is described as
\begin{widetext}
\begin{subequations}
\begin{align}~\label{DQMEFCS2_jcm}
 	d\hat{\rho}^\chi_{s,\rm R}(t)/dt=&-i[\hat{H},\hat{\rho}^\chi_{s,\rm R}]
 	+\mathcal{L}_{\chi_q}[\hat{\rho}^\chi_{s,\rm R}]+\sum_{\nu=1,2}\mathcal{L}_{\chi_p,\nu}[\hat{\rho}^\chi_{s,\rm R}],\\
\mathcal{L}_{\chi_q}[\hat{\rho}^\chi_{s,\rm R}]=&
[\hat{A}_q\hat{\rho}^\chi_{s,\rm R}\hat{D}^\dag_{q,-}(\chi_q)+\hat{A}^\dag_q\hat{\rho}^\chi_{s,\rm R}\hat{D}_{q,+}(\chi_q)+\hat{D}_{q,-}(-\chi_q)\hat{\rho}^\chi_{s,\rm R}\hat{A}^\dag_q+\hat{D}^\dag_{q,+}(-\chi_q)\hat{\rho}^\chi_{s,\rm R}\hat{A}_q]\nonumber\\
&-[(\hat{A}_q\hat{D}^\dag_{q,+}\hat{\rho}^\chi_{s,\rm R}+\hat{A}^\dag_q\hat{D}_{q,-}\hat{\rho}^\chi_{s,\rm R})+(\hat{\rho}^\chi_{s,\rm R}\hat{D}_{q,+}\hat{A}^\dag_q+\hat{\rho}^\chi_{s,\rm R}\hat{D}^\dag_{q,-}\hat{A}_q)],\\
\mathcal{L}_{\chi_p,\nu}[\hat{\rho}^\chi_{s,\rm R}]=&
[\hat{A}_{p,\nu}\hat{\rho}^\chi_{s,\rm R}\hat{D}^\dag_{p,\nu,-}(\chi_p)+\hat{A}^\dag_{p,\nu}\hat{\rho}^\chi_{s,\rm R}\hat{D}_{p,\nu,+}(\chi_p)+\hat{D}_{p,\nu,-}(-\chi_p)\hat{\rho}^\chi_{s,\rm R}\hat{A}^\dag_{p,\nu}+\hat{D}^\dag_{p,\nu,+}(-\chi_p)\hat{\rho}^\chi_{s,\rm R}\hat{A}_{p,\nu}]\nonumber\\
&-[(\hat{A}_{p,\nu}\hat{D}^\dag_{p,\nu,+}\hat{\rho}^\chi_{s,\rm R}+\hat{A}^\dag_{p,\nu}\hat{D}_{p,\nu,-}\hat{\rho}^\chi_{s,\rm R})+(\hat{\rho}^\chi_{s,\rm R}\hat{D}_{p,\nu,+}\hat{A}^\dag_{p,\nu}+\hat{\rho}^\chi_{s,\rm R}\hat{D}^\dag_{p,\nu,-}\hat{A}_{p,\nu})],
\end{align}
\end{subequations}
\end{widetext}
where $A^{nm}_{q}=\langle\phi_n|\hat{A}_{q}|\phi_m\rangle$and$A^{nm}_{p,\nu}=\langle\phi_n|\hat{A}_{p,\nu}|\phi_m\rangle$.
The corresponding system dissipative operators are
\begin{subequations}
 	\begin{align}
\hat{D}_{q,+}(\chi_q)&=\sum_{n,m}\frac{\gamma_q(\omega_d+E_{mn})}{2}n_q(\omega_d+E_{mn})\nonumber\\*&\quad\times e^{-i(\omega_d+E_{mn})\chi_q}A^{nm}_{q}|\varphi_n{\rangle}{\langle}\varphi_m|,~\label{Qpchi}\\
\hat{D}_{q,-}(\chi_q)&=\sum_{n,m}\frac{\gamma_q(\omega_d+E_{mn})}{2}[1+n_q(\omega_d+E_{mn})]\nonumber\\*&\quad\times e^{-i(\omega_d+E_{mn})\chi_q}A^{nm}_{q}|\varphi_n{\rangle}{\langle}\varphi_m|~\label{Qmchi}.\\
\hat{D}_{p,\nu,+}(\chi_p)&=\sum_{n,m}\frac{\gamma_p(\nu\omega_d+E_{mn})}{2}n_p(\nu\omega_d+E_{mn})\nonumber\\*&\quad\times e^{-i(\nu\omega_d+E_{mn})\chi_p}A^{nm}_{p,\nu}|\varphi_n{\rangle}{\langle}\varphi_m|,~\label{Ppchi_jcm}\\
\hat{D}_{p,\nu,-}(\chi_p)&=\sum_{n,m}\frac{\gamma_p(\nu\omega_d+E_{mn})}{2}[1+n_p(\nu\omega_d+E_{mn})]\nonumber\\*&\quad\times e^{-i(\nu\omega_d+E_{mn})\chi_p}A^{nm}_{p,\nu}|\varphi_n{\rangle}{\langle}\varphi_m|~\label{Pmchi_jcm}.
\end{align}
\end{subequations}
For the photon channel $p$ and the two-level-system channel $q$, the system transition operators are $\hat A_{p,1}=\hat a$, $\hat A_{p,2}=\hat a^{2}$, and $\hat A_{q}=\hat\sigma_{-}$, respectively.
In particular, the two-photon dissipative channel involves the
transition matrix elements of $\hat{a}^{2}$ and the
system-reservoir exchange frequency $2\omega_d+E_{mn}$.
The different dissipative channels are assumed to be independent,
with negligible environmental cross correlations.
The total dissipative superoperator can therefore be written as the
sum of the individual channel contributions.

\section{Floquet master equation with full counting statistics}\label{app:floquet}

Full counting statistics is considered as one powerful method to measure the currents and current fluctuations in nonequilibrium systems~\cite{esposito2009rmp}.
Here by including the counting fields into the Hamiltonian
$\hat{H}_{\textrm{tot}}(t)=\hat{H}_{\textrm{DS}}(t)+\sum_\mu(\hat{H}_{b,\mu}+\hat{V}_{\mu})$,
we obtain the modified total Hamiltonian  as
\begin{eqnarray}
    \hat{H}^{\chi}_{\textrm{tot}}(t)=\hat{H}_{\textrm{DS}}(t)+\sum_\mu(\hat{H}_{b,\mu}+\hat{V}_{\chi_\mu}),
\end{eqnarray}
where
$V_{\chi_\mu}=\sum_{k}({g}_{k,\mu}\hat{b}^\dag_{k,\mu}\hat{A}_\mu e^{i\chi_\mu\omega_{k,\mu}/2}+H.c.)$,
with the counting fields $\chi=\{\chi_\mu\}$.
Considering weak system-reservoir interactions, we perturb $\hat{V}_{\chi_\mu}$. Under the Born-Markov approximation, the driven master equation combined with full counting statistics (e.g., counting the flow into the right reservoir) is expressed as
\begin{eqnarray}
\frac{d\rho^{\chi}_s(t)}{dt}&=&-i[H_{\rm DS}(t),\rho^{\chi}_s(t)]-\sum_\mu\int^\infty_0d\tau{\times}\\
&&\textrm{Tr}_b
\{[V_{\chi_\mu},[V_{\chi_\mu}(t-\tau,t),\rho^{\chi}_s(t){\otimes}\rho_b]_\chi]_\chi\},\nonumber
\end{eqnarray}
where $\rho^{\chi}_s(t)$ is the system density matrix embedded with full counting statistics,
$V_{\chi_\mu}(t-\tau,t)=U^\dag_0(t-\tau,t)V_{\chi_\mu}U_0(t-\tau,t)$,
$U_0(t^\prime,t)=U_0(t^\prime)U^\dag_0(t)$ and $U_0(t)=\mathcal{T}\exp[-i\int^t_0d\tau(H_{\rm {DS}}(\tau)+\sum_\mu H_{b,\mu})]$,
with the reservoir term $H_{b,\mu}=\sum_k\omega_{k,\mu}b^\dag_{k,\mu}b_{k,\mu}$,
 the reservoir equilibrium density operator  is given by $\rho_{b}{\propto}\exp(-\sum_\mu H_{b,\mu}/k_BT_\mu)$,
and the commutating relation is given $[A_\chi,B_\chi]_\chi=A_\chi{B}_\chi-B_\chi{A}_{-\chi}$.
Then in the Floquet basis i.e. $[H_{\rm DS}(t)-i\frac{d}{dt}]|\psi_\alpha(t){\rangle}=\varepsilon_\alpha|\psi_\alpha(t){\rangle}$
we consider the evolution time is much longer than the characteristic driving time $T=2\pi/\Omega$,
which leads to the dynamics of density matrix elements
\begin{widetext}
\begin{align}
\frac{d\rho^\chi_{\alpha\beta}(t)}{dt}
={}&-i(\varepsilon_\alpha-\varepsilon_\beta)
\rho^\chi_{\alpha\beta}(t)
+\frac{1}{2}
\sum_{m,\alpha',\beta',\mu}
\Bigg\{
\nonumber\\
& e^{i\Delta_{\beta'\beta,-m}\chi_\mu}
\Big[
\Gamma^-_\mu(\Delta_{\beta'\beta,-m})
\sigma^{\mu,-}_{\alpha\alpha',m}
\sigma^{\mu,+}_{\beta'\beta,-m}
+
\Gamma^+_\mu(-\Delta_{\beta'\beta,-m})
\sigma^{\mu,+}_{\alpha\alpha',m}
\sigma^{\mu,-}_{\beta'\beta,-m}
\Big]
\rho^\chi_{\alpha'\beta'}(t)
\nonumber\\
&+
e^{-i\Delta_{\alpha\alpha',m}\chi_\mu}
\Big[
\Gamma^-_\mu(-\Delta_{\alpha\alpha',m})
\sigma^{\mu,-}_{\alpha\alpha',m}
\sigma^{\mu,+}_{\beta'\beta,-m}
+
\Gamma^+_\mu(\Delta_{\alpha\alpha',m})
\sigma^{\mu,+}_{\alpha\alpha',m}
\sigma^{\mu,-}_{\beta'\beta,-m}
\Big]
\rho^\chi_{\alpha'\beta'}(t)
\nonumber\\
&-
\Big[
\Gamma^+_\mu(\Delta_{\alpha'\beta',-m})
\sigma^{\mu,-}_{\alpha\alpha',m}
\sigma^{\mu,+}_{\alpha'\beta',-m}
+
\Gamma^-_\mu(-\Delta_{\alpha'\beta',-m})
\sigma^{\mu,+}_{\alpha\alpha',m}
\sigma^{\mu,-}_{\alpha'\beta',-m}
\Big]
\rho^\chi_{\beta'\beta}(t)
\nonumber\\
&-
\Big[
\Gamma^+_\mu(-\Delta_{\alpha'\beta',m})
\sigma^{\mu,-}_{\alpha'\beta',m}
\sigma^{\mu,+}_{\beta'\beta,-m}
+
\Gamma^-_\mu(\Delta_{\alpha'\beta',m})
\sigma^{\mu,+}_{\alpha'\beta',m}
\sigma^{\mu,-}_{\beta'\beta,-m}
\Big]
\rho^\chi_{\alpha\alpha'}(t)
\Bigg\}.
\label{FMEFCS}
\end{align}
\end{widetext}

where the density matrix element denotes $\rho^\chi_{\alpha\beta}(t)={\langle}\psi_\alpha(t)|\hat{\rho}^\chi_s(t)|\psi_\beta(t){\rangle}$,
the energy gap is specified as $\Delta_{\alpha\beta,m}=\varepsilon_\alpha-\varepsilon_\beta+m\Omega$,
transition rates are
$\Gamma^+_\mu(\omega)=\theta(\omega)\gamma_\mu(\omega)n_\mu(\omega)$
and
$\Gamma^-_\mu(\omega)=\theta(\omega)\gamma_\mu(\omega)[1+n_\mu(\omega)]$,
with the Heviside step function $\theta(\omega{>}0)=1$
and
$\theta(\omega{\le}0)=0$,
and the transition coefficients are
\begin{subequations}
\begin{align}
\sigma^{\mu,+}_{\alpha\beta,m}=&\frac{1}{T}\int^T_0dte^{-im{\Omega}t}{\langle}\psi_\alpha(t)|\hat A^{\dag}_\mu|\psi_\beta(t){\rangle},\\
\sigma^{\mu,-}_{\alpha\beta,m}=&\frac{1}{T}\int^T_0dte^{-im{\Omega}t}{\langle}\psi_\alpha(t)|\hat A_\mu|\psi_\beta(t){\rangle}.    
\end{align}    
\end{subequations}
The transition coefficients come from the relation
${\langle}\psi_\alpha(t)|\hat{A}_\mu|\psi_\beta(t){\rangle}=\sum_me^{im{\Omega}t}\sigma^{\mu,-}_{\alpha\beta,m}$. And the Fourier component is specified as
\begin{eqnarray}
 \sigma^{\mu,-}_{\alpha\beta,m}
=\sum_n{\langle}\psi_{\alpha,n}|\hat A_\mu|\psi_{\beta,n-m}{\rangle},
\end{eqnarray}
with $|\psi_{\alpha}(t){\rangle}=\sum_ne^{-in{\Omega}t}|\psi_{\alpha,n}{\rangle}$.

\end{document}